\documentclass[12pt]{article}
\usepackage{amsmath}
\usepackage{lineno}
\usepackage{epsfig} 
\usepackage{hhline}
\usepackage{amssymb}
\usepackage{times}
\usepackage{cite}
\usepackage{array}
\usepackage{multirow}
\usepackage{rotating}

\usepackage{color}
\definecolor{rltred}{rgb}{0.75,0,0}
\definecolor{rltgreen}{rgb}{0,0.5,0}
\definecolor{rltblue}{rgb}{0,0,0.75}

\newif\ifpdf
    \pdffalse          

\ifpdf
\usepackage{thumbpdf}
\usepackage[pdftex,
        colorlinks=true,
        urlcolor=rltblue,       
        filecolor=rltgreen,     
        linkcolor=rltred,       
        pdftitle={Example File for pdflatex},
        pdfauthor={H1},
        pdfsubject={Template file},
        pdfkeywords={High-Energy Physics, Particle Physics},
        pdfpagemode=None,
        bookmarksopen=true]{hyperref}

\fi

\newlength{\dinwidth}
\newlength{\dinmargin}
\newcommand{\gev}{\rm GeV}

\def\ra{\rightarrow}

\def\be{\begin{equation}}
\def\ee{\end{equation}}
\def\bea{\begin{eqnarray}}
\def\eea{\end{eqnarray}}

\begin{document}

\pagestyle{empty}
\begin{titlepage}
\noindent
DESY 06-240 \\
December 2006
\vspace*{2cm}

\begin{center}
  \Large
    {\bf Production of {\boldmath $D^{* {\mathbf \pm}}$} Mesons
with Dijets in Deep-Inelastic Scattering at HERA}

  \vspace*{1cm}
    {\Large H1 Collaboration}
\end{center}

\begin{abstract}

\noindent 
Inclusive $D^{*\pm}$ production is measured in deep-inelastic $ep$ 
scattering at HERA with the H1 detector. In addition, the production of dijets in events with a $D^{*\pm}$ meson is investigated. The analysis covers values of photon virtuality 
$2 \le Q^2 \le 100 \ \mbox{GeV}^2$ and of inelasticity
$0.05 \le y \le 0.7$. Differential cross sections are measured as a function of $Q^2$ and $x$ and of various $D^{*\pm}$ meson and jet
observables. Within the experimental and theoretical uncertainties all measured cross sections are found to be adequately described by 
next-to-leading order (NLO) QCD calculations, based on the photon-gluon fusion process and DGLAP evolution, without the need for an additional resolved component of the photon beyond what is included at NLO. A reasonable description of the data is also achieved by a prediction based on the CCFM evolution of partons involving the $k_{\rm T}$-unintegrated gluon distribution of the proton.

\end{abstract}

\vspace{1.5cm}

\begin{center}
To be submitted to Eur. Phys. J. {\bf C}
\end{center}

\end{titlepage}

\begin{flushleft}
A.~Aktas$^{10}$,               
V.~Andreev$^{24}$,             
T.~Anthonis$^{4}$,             
B.~Antunovic$^{25}$,           
S.~Aplin$^{10}$,               
A.~Asmone$^{32}$,              
A.~Astvatsatourov$^{4}$,       
A.~Babaev$^{23, \dagger}$,     
S.~Backovic$^{29}$,            
A.~Baghdasaryan$^{37}$
% ...

\bigskip{\it
% ...
$ ^{4}$ Inter-University Institute for High Energies ULB-VUB, Brussels;
          Universiteit Antwerpen, Antwerpen; Belgium$^{ c}$ \\
% ...
$ ^{10}$ DESY, Hamburg, Germany \\
% ...
$ ^{23}$ Institute for Theoretical and Experimental Physics,
          Moscow, Russia$^{ k}$ \\
 $ ^{24}$ Lebedev Physical Institute, Moscow, Russia$^{ e}$ \\
 $ ^{25}$ Max-Planck-Institut f\"ur Physik, M\"unchen, Germany \\
% ...
$ ^{29}$ Faculty of Science, University of Montenegro,
          Podgorica, Montenegro$^{ e}$ \\
% ...
$ ^{32}$ Dipartimento di Fisica Universit\`a di Roma Tre
          and INFN Roma~3, Roma, Italy \\
% ...
$ ^{37}$ Yerevan Physics Institute, Yerevan, Armenia \\
% ...
$ ^{\dagger}$ Deceased \\

\bigskip
% ...
$ ^c$ Supported by FNRS-FWO-Vlaanderen, IISN-IIKW and IWT
      and  by Interuniversity
Attraction Poles Programme,
      Belgian Science Policy \\
% ...
$ ^e$ Supported by the Deutsche Forschungsgemeinschaft \\
% ...
$ ^k$ Partially Supported by Russian Foundation
      for Basic Research,  grants  03-02-17291
      and  04-02-16445 \\
% ...
}
\end{flushleft}

\newpage

\pagestyle{plain}

\section{Introduction}

Charm quark production in deep-inelastic $ep$ collisions at HERA is
of particular interest for testing calculations in the framework of perturbative quantum chromodynamics (pQCD). This process has the special feature of two hard scales: the photon virtuality $Q^2$ and the charm quark mass. In the case of jet production the transverse energy of the jets provides a further hard scale. In leading order (LO) QCD, the photon-gluon fusion process $\gamma g \ra c \bar c\ $ is the dominant production mechanism.

Results on inclusive $D^{*\pm}$ meson production in deep-inelastic scattering (DIS) have been published by the H1 and ZEUS collaborations
\cite{DESY01100,DESY03115,DESY99101,h1gluon,h1f2c,zeusf2c,h1dfrac}.
The analysis described in this paper uses data collected during 1999 and 2000, corresponding to a larger integrated luminosity of 
47~$\mbox{pb}^{-1}$ than used in previous H1 publications 
\cite{DESY01100,h1gluon,h1f2c}. As a result, the production of inclusive $D^{*\pm}$ mesons is measured in DIS with increased precision. 

The mechanism of charm production is further explored by studying the production of dijets in events with a $D^{*\pm}$ meson for the first time in DIS. This study is referred to in the following as the analysis of ``$D^{*\pm}$ mesons with dijets''.
One of the jets typically contains the $D^{*\pm}$ meson. According to Monte Carlo simulation studies the jet containing the $D^{*\pm}$ meson provides a very good approximation of the kinematics of the associated charm quark. The other jet usually also gives a good approximation of the energy and direction of the second charm quark or of a radiated gluon. Regarding the theoretical predictions, the measurement of jets has a reduced sensitivity to fragmentation uncertainties compared to the method based solely on measuring the $D^{*\pm}$ meson. Thus it corresponds more closely to a measurement of the underlying partons and is therefore expected to lead to more reliable theoretical predictions.

The photon-gluon fusion process of charm quark pair production provides sensitivity to the gluon distribution in the proton. The dijet event sample is used to measure the observed gluon momentum fraction $x_{\rm g}^{\rm obs}$. The azimuthal angular correlation 
$\Delta\phi$ between the two leading jets is investigated because of its sensitivity to initial state gluon emissions. These measurements are compared with QCD calculations based on either collinear or $k_{\rm T}$-factorisation and using gluon densities obtained from QCD fits to HERA inclusive DIS data. 
The photon, in addition to its coupling as a point-like object in the hard scattering process, exhibits a partonic structure \cite{g1} which can be resolved by the hard scale present in the process, and which is described by a photon structure function. Dijets are used to measure the observed fractional momentum of the parton from the photon taking part in the hard interaction. Comparing this distribution with pQCD calculations allows a test for resolved contributions of the photon which go beyond what is already included in calculations of the photon-gluon fusion process at 
next-to-leading order (NLO).

In this paper measurements of single and double differential cross sections for the production of $D^{*\pm}$ mesons and $D^{*\pm}$ mesons with dijets are reported. They are compared to perturbative QCD calculations using different implementations of the evolution of the gluon from the proton.

\section{QCD Models}

QCD models for data corrections and for comparison with measured 
cross sections are introduced in the following two sections. Relevant parameters used in the Monte Carlo (MC) simulations and the NLO calculations are described and summarized in table~\ref{tabsettings}.

\subsection{QCD Models for Data Corrections}
Monte Carlo models are used to generate charm events and to simulate
detector effects in order to determine the acceptance and the
efficiency for selecting events with a $D^{*\pm}$ meson only and with
dijets and to estimate the systematic uncertainties associated with the measurements. All the events are passed through a detailed simulation of the detector response based on the GEANT simulation program~\cite{geant} and are reconstructed using the same reconstruction software as used for the data.

The Monte Carlo programs  RAPGAP~\cite{rapgap} and 
HERWIG~\cite{herwig} are used to generate in DIS the direct process of photon-gluon fusion to a heavy (charm or beauty) quark anti-quark pair, where the photon acts as a point-like object. In addition, they allow the simulation of charm production via resolved processes, where the photon fluctuates into partons, one of which interacts with a parton in the proton and the rest produces the photon remnant. Both programs use LO matrix elements with massive (massless) charm quarks for the direct (resolved) processes. Parton showers based on DGLAP evolution are used to model higher order QCD effects. The masses of the $c$ and $b$ quarks are set to $m_c=1.5\;\gev$ and $m_b=4.75\;\gev$. In RAPGAP the hadronization of partons is performed using the Lund String model as implemented in PYTHIA \cite{pythia}. For the longitudinal fragmentation of the charm quark into the $D^{*\pm}$ meson the Bowler parametrisation~\cite{bowler} is taken with parameters as obtained by BELLE \cite{belle}. The fragmentation fraction $f(c\ra D^{*+})=0.257\pm0.015\pm0.008$ \cite{hadro} is used.
RAPGAP is interfaced to HERACLES \cite{heracles} in order to simulate the radiation of a photon from the incoming or outgoing lepton including virtual effects. The simulation of such effects is only available for direct processes. For resolved processes, similar QED effects are assumed. RAPGAP is normally used for the determination of the detector acceptance and efficiency. The effect of a different model on the detector acceptance and efficiency is investigated by using the HERWIG program, which is based on the cluster hadronization model. For both models small differences in the spectrum of the transverse momentum of the $D^{*\pm}$ meson in comparison with the data are corrected by reweighting the spectrum to that observed in the data.

\subsection{QCD Models used for Comparing with Data}

In this paper the experimental results are compared with predictions considering three active flavours ($u$, $d$, $s$) in the proton
(fixed-flavour-number scheme FFNS) and massive charm quarks produced
via photon-gluon fusion. The results are also compared with a
calculation based on the zero-mass variable-flavour-number scheme 
(ZM-VFNS), where the charm quark occurs also as an incoming parton and is treated as massless. In the case of the FFNS two different pQCD approaches are used: an NLO calculation 
\cite{riemersma,riemersma2,harris} based on the conventional collinear factorisation and the DGLAP evolution equations \cite{dglapref} and another prediction based on $k_{\rm T}$-factorisation and parton evolution according to the CCFM equations \cite{ccfm}. A beauty contribution of $1.5 \pm 0.5$ times the QCD prediction is added to the charm expectations of the calculations with massive charm quarks to encompass the range of beauty cross section measurements 
\cite{hera_beauty}.
According to Monte Carlo studies, the contribution of beauty quarks is approximately $3\%$ for events with $D^{*\pm}$ mesons and $7\%$ for events with $D^{*\pm}$ mesons with dijets. 
The basic parameter choices for the various pQCD programs and the range of their variations are summarised in table~\ref{tabsettings}. Each of the variations is performed independently to determine specific cross section uncertainties. These uncertainties and the assumed error on the beauty contribution are added in quadrature to obtain the total theoretical uncertainty which is shown in the figures as a band. The programs implementing the above approaches are discussed in the following.

\begin{table}[htb]
\begin{center}
\footnotesize
{\begin{tabular}{l@{\hspace{.4cm}}c@{\hspace{.4cm}}c@{\hspace{.4cm}}c@{\hspace{.4cm}}c@{\hspace{.4cm}}}
  & {\bf $p$}-PDF
  & {\bf $\mu$}
  & {\bf $m_c$} [GeV]
  & {fragmentation} \\[1mm]
  & {\bf $\gamma$}-PDF & & & \\[1mm]
\hline \hline \\
RAPGAP
  & CTEQ6L~\cite{cteq6} \footnotemark
  & $\sqrt{Q^2 +p_{\rm T}^2}$
  & 1.5
  & Bowler a=0.22, b=0.56 \cite{belle} \\[1mm]
  & SAS-G 2D \cite{sasg_v2} & & & \\[1mm]
\hline \\
CASCADE
  & A0~\cite{hancassets}
  & $\mu_{\rm r}=\sqrt{4m_c^2+p_{\rm T}^2}$
  & 1.5
  & Bowler a=0.22, b=0.56 \cite{belle} \\[1mm]
\hspace*{5mm} variation
  & A+, A-~\cite{hancassets}
  & 1/2 $\mu_{\rm r}$ -- 2 $\mu_{\rm r}$
  & 1.4 -- 1.6
  & Peterson $\epsilon_c$: 0.025 -- 0.060 \cite{suzanna,zeusfrag} \\[1mm]
\hline \\
HVQDIS
  & CTEQ5F3~\cite{cteq5f3}
  & $\sqrt{Q^2 +4m_c^2}$
  & 1.5
  & Kartvelishvili $\alpha = 3.0$~\cite{suzanna} \\[1mm]
\hspace*{5mm} variation
  & {}
  & $\text{max\,}($2 $m_c$, 1/$\sqrt 2$ $\mu)$ -- 
  & 1.4 -- 1.6
  & $\alpha$: 2.5 -- 3.5 \\[1mm]
\hspace*{5mm}
  & {}
  & $\sqrt 2$ $\mu$~\cite{hvqdis}
  & {}
  & {} \\[1mm]
\hline \\
ZM-VFNS
  & CTEQ6.1M~\cite{cteq6}
  & $\sqrt{(Q^2 +(p^*_{\rm T})^2)/2}$
  & 1.5
  & ~\cite{zmfrag} \\[1mm]
\hspace*{5mm} variation
  & {}
  & 1/$\sqrt 2$ $\mu$ -- $\sqrt 2$ $\mu$
  & {}
  & {} \\[1mm]
\hline
\hline
\end{tabular}
\caption{\label{tabsettings}
Parton density functions (PDFs) and parameters used in the Monte Carlo simulations and the NLO programs. The renormalisation and factorisation scales are set equal,
$\mu = \mu_{\rm r} =\mu_{\rm f}$ (apart from CASCADE where
$\mu_{\rm f}$ is kept fixed at 
$\mu_{\rm f}=\sqrt{\hat{s} + Q^2_{\rm T}}$, 
where the invariant mass squared and the transverse momentum squared of the $c\bar{c}$-pair are denoted by $\hat{s}$ and $Q_{\rm T}^2$, respectively), $m_c$ is the charm quark mass and $\epsilon_c$ and 
$\alpha$ are the fragmentation parameters according to the Peterson and the Kartvelishvili parametrisations. The range of variation for the different parameters is also indicated. Each of the variations is performed independently.}}
\end{center}
\end{table}
\footnotetext{ For data corrections the CTEQ5L\cite{cteq5f3} parton
density function is used for the proton.}

The NLO $\cal O$($\alpha_s^2$) QCD FFNS predictions are calculated using the program HVQDIS. The CTEQ5F3~\cite{cteq5f3} parton densities of the proton are used. Charm quarks are fragmented in the $\gamma p$ center-of-mass frame into $D^{*\pm}$ mesons using the Kartvelishvili {\it et al.}~\cite{kartvelishvili} parametrisation for the fragmentation function which, for HVQDIS, yields a better description of the H1 data \cite{suzanna} than the Peterson~\cite{peterson} parametrisation. ``Decays'' of beauty quarks to $D^{*\pm}$ mesons are parametrised by adapting the longitudinal as well as the transverse fragmentation distribution from RAPGAP using the Peterson model with $\epsilon_b=0.0080$. 

In order to compare parton level dijets of HVQDIS with hadron level
dijets of the data, hadronization corrections have to be applied to
the NLO calculations. They are estimated by using the RAPGAP and
HERWIG Monte Carlo models described in the previous section. Dijets are reconstructed at the parton level from the generated quarks and gluons after the parton showering step, using the same jet algorithm and selection cuts as at the hadron and detector levels. For each
kinematic bin the ratio of the hadron to parton level cross section is
calculated. The average values from the two Monte Carlo models are taken as hadronization corrections to the NLO predictions. They
vary typically between $-5\%$ and $-20\%$ and occasionally amount to
$-40\%$ and up to $+40\%$. 
Their uncertainty is taken to be half the difference between the predictions of the two models. This uncertainty is added in quadrature to the other theoretical uncertainties to obtain the total error for the HVQDIS prediction of $D^{*\pm}$ mesons with dijets.

The predictions based on the CCFM evolution equation are calculated using the CASCADE~\cite{cascade} program. In CASCADE the direct process $\gamma g \ra c \bar{c}$ is implemented using off-shell matrix elements convoluted with the $k_{\rm T}$-unintegrated gluon distribution of the proton. The parametrisation 
set-A0~\cite{hancassets} is used for the latter. It has been determined from a fit to $F_2$ data published by H1~\cite{h1f2jung} and ZEUS~\cite{zeusf2jung}. Time-like parton showers off the charm quark and anti-quark but not off initial state gluons are implemented. The hadronization of partons is performed in the same way as described in section~2.1 for RAPGAP. The sensitivity of the cross section to the parametrisation used for the longitudinal fragmentation of the charm quark into the $D^{*\pm}$ meson is investigated by using the Peterson function with parameters as obtained by HERA measurements \cite{suzanna,zeusfrag} instead of the Bowler function. Resolved photon processes are not implemented in CASCADE.

An NLO QCD calculation of inclusive hadron production in DIS in the ZM-VFNS has recently become available \cite{massless}. This calculation treats the charm quark as massless. The contribution of the partonic subprocesses $\gamma q \ra q g$ and
$\gamma g \ra q \bar{q}$ for the production of charm and beauty quarks is considered at LO. The NLO corrections are large in certain kinematic regions. To allow a comparison of the data to the predictions from the ``massless'' calculation, it is required that the $D^{*\pm}$ meson has a transverse momentum $p_{\rm T}^* >2$~GeV in the
$\gamma p$ center-of-mass frame. At present, predictions from the ``massless" approach exist only for inclusive $D^{* \pm}$ meson production and not for the production of $D^{* \pm}$ mesons with dijets.

\section{H1 Detector}

The data presented were collected with the H1 detector at HERA in the years 1999 and 2000. During this period HERA operated with 27.5 GeV positrons and 920 GeV protons colliding at a center-of-mass energy of $\sqrt{s} = 318\ \mbox{GeV}$. The data sample used for this analysis amounts to an integrated luminosity of 
${\cal L} = 47.0\ \mbox{pb}^{-1}$.

A detailed description of the H1 detector is given in \cite{h1det}. Here only the relevant components for this analysis are described. The positive $z$-axis of the H1 reference frame, which defines the forward direction, is given by the proton-beam direction. The scattered positron is identified and measured in the SpaCal calorimeter~\cite{spacal}, a lead-scintillating fibre calorimeter situated in the backward region of the H1 detector, covering the polar angular range $153 ^\circ < \theta < 177.8 ^\circ $. The SpaCal also provides time-of-flight information which is used for triggering purposes. Hits in the backward drift chamber (BDC) are used to improve the identification of the scattered positron and the measurement of its angle. Charged particles emerging from the interaction region are measured by the Central Tracking Detector (CTD), which covers a range 
$20^\circ< \theta<160^\circ$. The CTD comprises two large cylindrical Central Jet drift Chambers (CJCs) and two $z$-chambers situated concentrically around the beam-line within a solenoidal magnetic field of $1.15~{\rm T}$\@. It also provides triggering information based on track segments measured in the $r$-$\phi$-plane of the
CJCs, and on the $z$-position of the event vertex obtained from the double layers of two multi-wire proportional chambers (MWPCs).
In the central and forward region the track detectors are surrounded by a finely segmented Liquid Argon Calorimeter (LAr). It consists of an electromagnetic section with lead absorbers and a hadronic section with steel absorbers and covers the range 
$4 ^\circ < \theta < 154 ^\circ $.

The luminosity determination is based on the measurement of the                                                         Bethe-Heitler process $(ep \ra ep\gamma)$, where the photon is detected in a calorimeter located downstream of the interaction      point in the positron beam direction.

\section{Event Selection}

At fixed center-of-mass energy, $\sqrt{s}$, the kinematics of the
inclusive scattering process $ep \ra eX$ is determined by any two of the following Lorentz-invariant variables: the Bjorken scaling variable $x$, the inelasticity $y$, the square of the four-momentum-transfer $Q^2$ and the invariant mass squared $W^2$ of the hadronic final state. In this analysis these variables are determined from the measurement of the scattered positron energy, $E^\prime_e$, and its polar angle, $\theta^\prime_e$, according to
\begin{linenomath}\begin{equation}
\begin{array}{ccc}\displaystyle
Q^2=4E_eE^\prime_e\cos^2\left(
\frac{\theta^\prime_e}{2}\right)&\quad\quad&\displaystyle y=1-\frac{
E^\prime_e}{ E_e} \sin^2\left(\frac{ \theta^\prime_e}{2}\right)
\cr\cr\displaystyle x=\frac{ Q^2}{ ys}&\quad\quad&\displaystyle
W^2=Q^2\left(\frac{ 1-x}{ x}\right),\cr
\end{array}
\end{equation}\end{linenomath}
where $E_e$ is the incident positron beam energy.

The analysis covers the kinematic region 
$2 \le Q^2 \le 100 \ \mbox{GeV}^2$ and $0.05 \le y \le 0.7$.
DIS events were triggered by requiring signals from the central drift chambers and the multi-wire proportional chambers in coincidence with                                                                               signals from the scattered positron in the SpaCal. The identification and selection of the scattered positron is performed as described in \cite{DESY01100}.

$D^{*\pm}$ mesons are reconstructed using the decay chain 
$D^{*+}\ra D^0\pi^+_s\ra K^-\pi^+\pi^+_s$ (and c.c.), where the notation $\pi_s$ is used for the slow pion. The three decay tracks are measured in the central track detector. For all tracks particle identification is applied using the measurement of the energy loss, ${\rm d}E/{\rm d}x$, in the CJCs. The invariant mass of the 
$K^{-}\pi^{+}$ system is required to be consistent with the nominal $D^0$ mass within two standard deviations. The signal is extracted from a simultaneous fit to the distribution of 
$\Delta m = m_{K \pi \pi} - m_{K \pi}$ of the $D^{* \pm}$ meson
candidates and of the wrong sign combinations
$(K^{\pm} \pi^{\pm}) \pi^{\mp}_s$ which provide a good description of the shape of the uncorrelated background. Further details are
described in \cite{sebthesis,DESY01100,h1f2c}. The range of the
transverse momentum and the pseudorapidity of the $D^{* \pm}$ meson
is restricted to $1.5 \le p_{\rm T} \le 15\ \mbox{GeV}$ and
$|\eta| \le 1.5$, where $p_{\rm T}$ and $\eta$ are defined in the
laboratory frame with $\eta=-\ln\tan\left(\frac{\theta}{2}\right)$.
From the fit a total of $2604\pm 77$ $D^{* \pm}$ mesons is obtained.

In order to define the ``$D^{*\pm}$ meson with dijets'' sample, the 
$k_{\rm T}$-cluster algorithm~\cite{ktcluster_alg} in its inclusive mode is applied to the hadronic final state objects in the Breit frame
for events containing a $D^{*\pm}$ meson candidate. The hadronic objects are built by combining the energy depositions in the SpaCal and the LAr calorimeter with the track momenta measured in the tracking system. Such objects have improved energy and angular resolution and are well suited for the low transverse momentum jets produced in charm events at HERA. For jets with transverse energy 
$3\ \le E_{\rm T}^{\rm jet} \le \ 5$~GeV the energy resolution is about $20\%$. When applying the jet algorithm, the four-vector of the
reconstructed  $D^{* \pm}$ meson is used instead of the four-vectors of its three decay particles. The jet algorithm is used with the separation parameter set to unity and using the $E$-recombination scheme, in which the four-vectors of the hadronic objects are added.

For the dijet selection, the transverse energies of the leading jets in the Breit frame are required to be 
$E_{\rm T}^{\rm jet 1(2)} \ge 4(3)$~GeV, and their pseudorapidities
in the laboratory frame have to fulfill 
$-1 \le \eta_{\rm lab}^{\rm jet 1,2} \le 2.5$. Down to these low jet transverse energies Monte Carlo simulation studies show a very good correlation of the parton and reconstructed jet quantities for both jets. From a fit to the $\Delta m$ distribution a total of $668 \pm 49$ $D^{*\pm}$ mesons is obtained for events fulfilling the dijet requirements. In about $90 \%$ of the events the $D^{*\pm}$ meson belongs to one of the two leading jets in agreement with Monte Carlo predictions.

\section{Cross Section Determination and Systematic Errors}

The total visible cross section for inclusive $D^{*\pm}$ meson production in deep-inelastic $ep$ scattering, including requirements on the DIS phase space and on the kinematics of the $D^{*\pm}$ meson, is calculated from the observed number of $D^{*\pm}$ mesons, 
$N_{D^{*\pm}}$, according to
\begin{linenomath}\begin{equation}\label{eqsigma}
  \sigma(e^+ p \ra e^+ D^{*\pm} X) = \frac{N_{D^{*\pm}}
  }{{\cal L} \cdot Br \cdot \epsilon
  \cdot(1+\delta_{\text{rad}})}.
\end{equation}\end{linenomath}
Here $Br$ refers to the branching ratio
$Br(D^{*+} \ra D^0 \pi^+) \cdot Br(D^0\ra K^- \pi^+) = 
0.0258$~\cite{branching} and ${\cal L}$ to the integrated luminosity.
The factor $\epsilon=31\%$ corrects for the acceptance loss due to the track selection cuts and the detector efficiency and resolution. The QED radiative correction, $\delta_\text{rad}$, amounts to $-2\%$. For differential cross sections the data sample is divided into bins and the number of $D^{*\pm}$ mesons is extracted in each bin separately. The visible total and differential cross section for the $D^{*\pm}$ mesons with dijets is defined in a similar way.

The systematic errors on the cross section measurements are estimated as follows (numbers are given for the total visible cross section):
\begin{itemize}
\item
The trigger efficiency is monitored using data samples with independent trigger conditions. Its associated uncertainty is estimated to be 
$2\%$. 
\item
The uncertainty in the track reconstruction efficiency
leads to an error of $4.8\%$.
\item
To account for possible imperfections in the description of low transverse momentum tracks by the Monte Carlo simulation the requirement for the minimal transverse momentum of the $\pi_s$ candidate is varied from $120\ \mbox{MeV}$ to $150\ \mbox{MeV}$.
This leads to a cross section change of $4\%$.
\item
An error of $3\%$ due to the uncertainty on the d$E$/d$x$ measurement used for particle identification is estimated.
\item
The systematic error of the $D^{*\pm}$ signal extraction procedure is estimated to be $4.9\%$.
\item
The estimated uncertainties on the measurement of the scattered positron energy $E^\prime_e$ of $1\%$ and of its polar angle 
$\theta^\prime_e$ of 1~mrad lead to an error of $1.8\%$.
\item
A model uncertainty of $1\%$ is estimated from the difference in the correction factor for the acceptance and efficiency obtained with
RAPGAP and HERWIG.
\item
An uncertainty of $1.5\%$ is caused by the error in the determination of the luminosity.
\item
The uncertainty on the $D^{*+}$ and $D^0$ branching ratios contributes an error of $2.5\%$.
\item
For the analysis of $D^{*\pm}$ mesons with dijets additional errors of $4\%$ and $1\%$ are taken into account. They arise from the uncertainty in the hadronic energy scale of the LAr ($4\%$) and the
SpaCal ($7\%$), respectively.
\end{itemize}
The systematic uncertainties on the total visible cross section are summarized in table~\ref{tabsys}. For the differential cross section measurements the systematic errors are evaluated separately for each bin. All contributions are added in quadrature to obtain the total systematic errors.

\begin{table}
\begin{center}
{\footnotesize
\begin{tabular}{l@{\hspace{.4cm}}c@{\hspace{.2cm}}r@{\hspace{.2cm}}c@{\hspace{.2cm}}}
Systematic Uncertainties
 & $D^{*\pm}$ meson &  & $D^{*\pm}$ + dijets \rule[-2mm]{0mm}{5mm}
\cr\noalign{\smallskip} \hline\hline\noalign{\smallskip} 
Trigger efficiency &  & $\pm$~$2\%$ &
\cr\noalign{\smallskip} Track reconstruction efficiency & 
& $\pm$~$4.8\%$ &
\cr\noalign{\smallskip} d$E$/d$x$ measurement & & $\pm$~$3\%$ &
\cr\noalign{\smallskip} Description of $\pi_s$ tracks with low 
$p_{\rm T}$ & & $-$~$4\%$ &
\cr\noalign{\smallskip} $D^{*\pm}$ signal extraction & 
& $\pm$~$4.9\%$ &
\cr\noalign{\smallskip} Measurement errors on $E^\prime_e$ and $\theta^\prime_e$ & & $\pm$~$1.8\%$ &
\cr\noalign{\smallskip} Luminosity measurement & & $\pm$~$1.5\%$ &
\cr\noalign{\smallskip} Branching ratio & & $\pm$~$2.5\%$ &
\cr\noalign{\smallskip} Model dependence of acceptance and reconstruction efficiency & & $\pm$~$1\%$ &
\cr\noalign{\smallskip} Hadronic energy scale of the LAr Calorimeter  & -- & & $\pm$~$4\%$
\cr\noalign{\smallskip} Hadronic energy scale of the SpaCal Calorimeter & -- & & $\pm$~$1\%$
\cr\noalign{\smallskip} \hline\noalign{\smallskip} &$+$~$8\%$& & $+$~$9\%$
\cr\noalign{\smallskip} &$-$~$9\%$& & $-$~$11\%$
\cr\noalign{\smallskip} \hline\hline\noalign{\smallskip}
\end{tabular}}
\caption{\label{tabsys}{Experimental systematic uncertainties on the total visible production cross section for inclusive $D^{*\pm}$ mesons and for $D^{*\pm}$ mesons with dijets.}}
\end{center}
\end{table}
\vspace*{-3mm}

\section{Inclusive $\mathbf D^{*\pm}$ Meson Cross Sections}

The cross section for inclusive $D^{*\pm}$ meson production in the DIS
kinematic region $2 \le Q^2 \le 100 \ \mbox{GeV}^2$, $0.05 \le y \le
0.7$ and in the visible $D^{*\pm}$ range $1.5 \le p_{{\rm T}}\le 15\
\mbox{GeV}$ and $|\eta| \le 1.5$ is found to be
\begin{linenomath}\begin{equation}\label{eqsigmad}
\sigma_{\rm vis}(e^+p \ra e^+D^{*\pm} X)=
6.99 \pm 0.20\, (\rm{stat.})\ ^{+ 0.57} _{-0.63}\, (\rm{syst.})\ \rm{nb}.
\end{equation}\end{linenomath}
A comparison with the predictions from HVQDIS and CASCADE is shown in
table~\ref{tabinclpred}. The models include a small beauty contribution as described in section~2. The predictions of both calculations are slightly below the data ($\sim 12\%$) but are consistent with the data within errors. The quoted theoretical errors include the variations of the scale $\mu$, the charm quark mass $m_c$ and the fragmentation parameters as indicated in 
table~\ref{tabsettings}. All these contributions to the theoretical uncertainty are of roughly similar size. These uncertainties and the assumed error on the beauty contribution, which leads to a relatively small cross section error, are all added in quadrature to define the total theoretical uncertainty. Use of the MRST2004F3NLO~\cite{mrst2004} instead of the CTEQ5F3 parton densities of the proton for HVQDIS results in a $9\%$ decrease of the cross section.
If for the $k_{\rm T}$-unintegrated gluon distribution of the proton in CASCADE the J2003 set-1~\cite{gluonj2003} parametrisation is used, instead of set-A0, the cross section increases by $3\%$.

\begin{table}[htb]
\begin{center}
{\footnotesize
\begin{tabular}{l@{\hspace{.4cm}}c@{\hspace{.4cm}}c@{\hspace{.4cm}}c@{\hspace{.4cm}}}
& H1 data & HVQDIS & CASCADE \rule[-1.2ex]{0pt}{0pt}  \\
\hline \hline \\
$\sigma_{\rm vis}(e^+p \ra e^+D^{*\pm} X)$ &
$6.99 \pm 0.20\, (\rm{stat.})\ ^{+ 0.57} _{-0.63}\, (\rm{syst.})\ \rm{nb}$
& $6.11 ^{+ 0.55} _{-0.61}$ nb & $6.19 ^{+ 0.72} _{-0.63}$ nb \rule[-2.2ex]{0pt}{0pt} \\
\hline \\
$\sigma_{\rm vis}(e^+p \ra e^+D^{*\pm}\text{jj} \ X)$ &
$1.60\pm 0.12\, (\rm{stat.})\ ^{+0.14} _{-0.18} \, (\rm{syst.})\ 
\rm{nb}$
& $1.35 ^{+0.17} _{-0.13}$ nb & $1.65 ^{+0.10} _{-0.09}$ nb \rule[-2.2ex]{0pt}{0pt} \\
\hline \\
${\frac{\textstyle \sigma_{\rm vis}(e^+p \ra e^+D^{*\pm} {\rm jj} \ X)}
{\textstyle \sigma_{\rm vis}(e^+p \ra e^+D^{*\pm} \ X)}}$ &
$0.228\pm 0.014\, (\rm{stat.}) \pm 0.011 \, (\rm{syst.})$ &
$0.221 ^{+0.029} _{-0.020}$ & $0.267 ^{+0.020} _{-0.022}$ \rule[-3.5ex]{0pt}{0pt}\\
\hline \hline
\end{tabular}
\caption{\label{tabinclpred}{Comparison of the cross sections for
inclusive $D^{*\pm}$ meson production and dijet production in
association with a $D^{*\pm}$ meson and of their ratio with the
predictions from HVQDIS and CASCADE. The errors on the predictions
include the variations of parameters, as indicated in 
table~\ref{tabsettings}, and the assumed $\pm 33\%$ error on the beauty contribution added in quadrature. }}}
\end{center}
\end{table}

\begin{figure}
\begin{center}
\vspace*{-6mm}
{\Large {\bf H1 ${\;\; \mathbf{ep \ra eD^{*\pm}X}}$}}
\vspace*{1mm}
\break\hspace*{-7mm}
\begin{minipage}{8.10cm}
\mbox{\epsfig{file=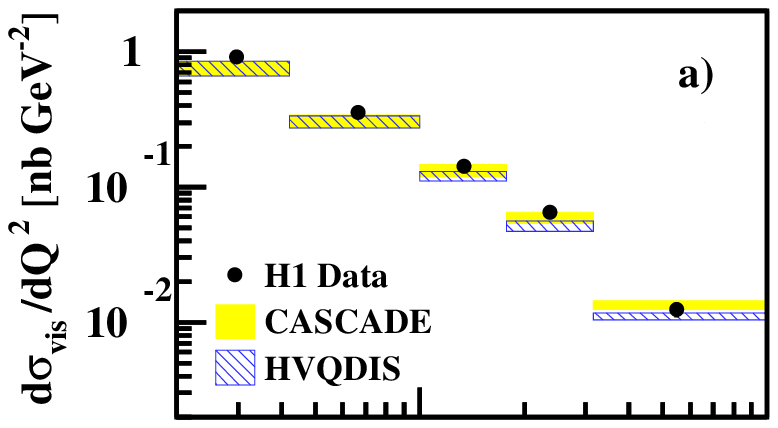,width=0.45\linewidth,bbllx=0,bblly=57,bburx=115,bbury=180}}\break
\mbox{\epsfig{file=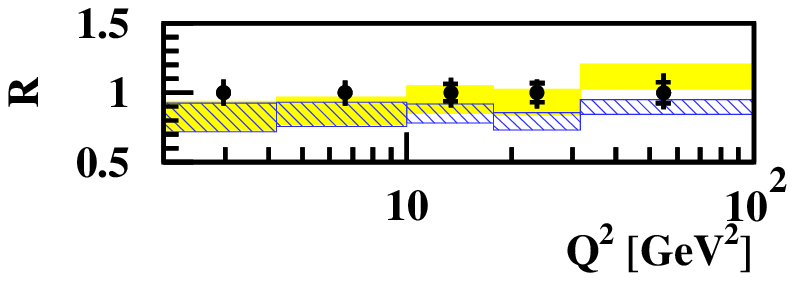,width=0.45\linewidth,bbllx=0,bblly=0,bburx=115,bbury=96}}
\end{minipage}
\begin{minipage}{8.10cm}
\mbox{\epsfig{file=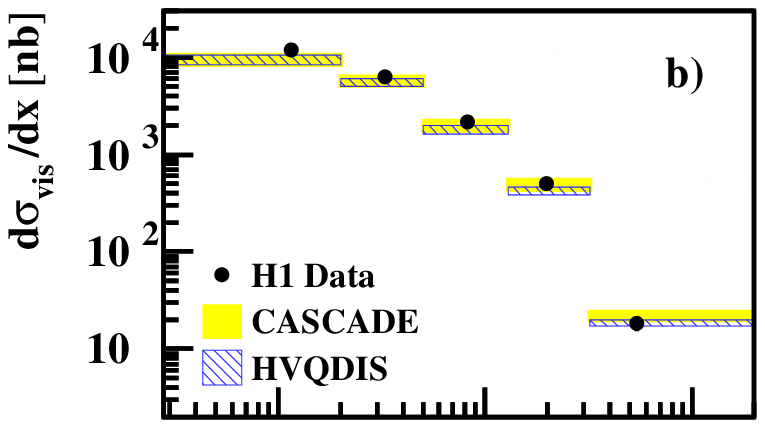,width=0.45\linewidth,bbllx=0,bblly=57,bburx=115,bbury=180}}\break
\mbox{\epsfig{file=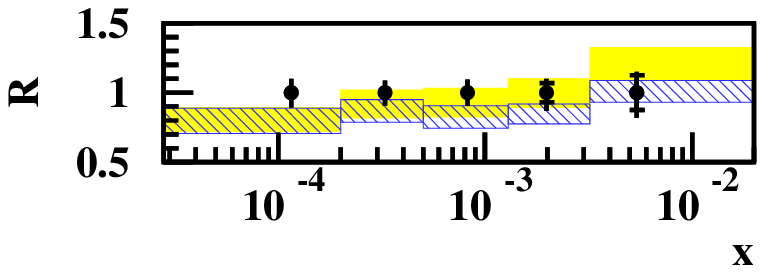,width=0.45\linewidth,bbllx=0,bblly=0,bburx=115,bbury=96}}
\end{minipage}
\vspace*{-6mm}
\break\hspace*{-7mm}
\begin{minipage}{8.10cm}
\mbox{\epsfig{file=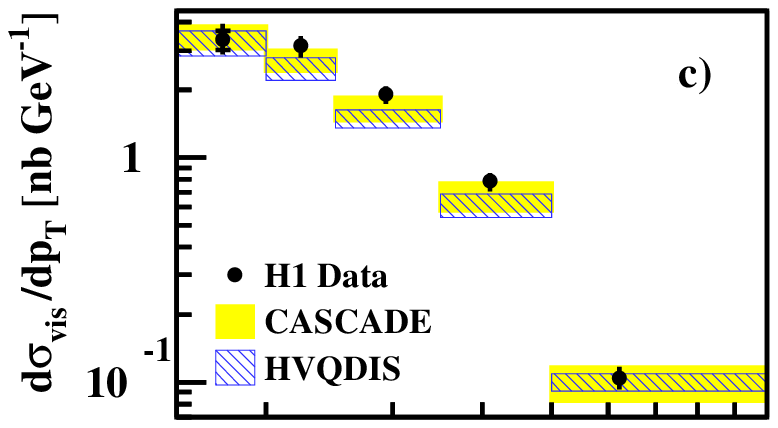,width=0.45\linewidth,bbllx=0,bblly=57,bburx=115,bbury=180}}\break
\mbox{\epsfig{file=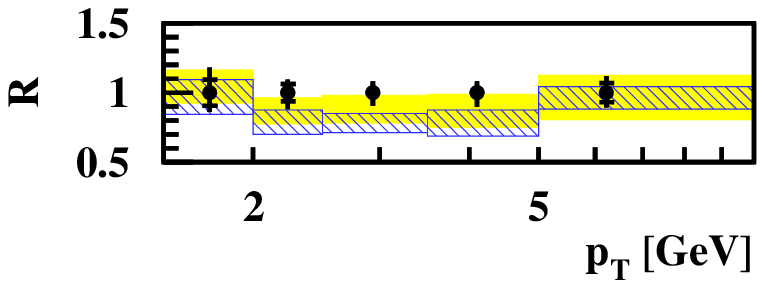,width=0.45\linewidth,bbllx=0,bblly=0,bburx=115,bbury=96}}
\end{minipage}
\begin{minipage}{8.10cm}
\mbox{\epsfig{file=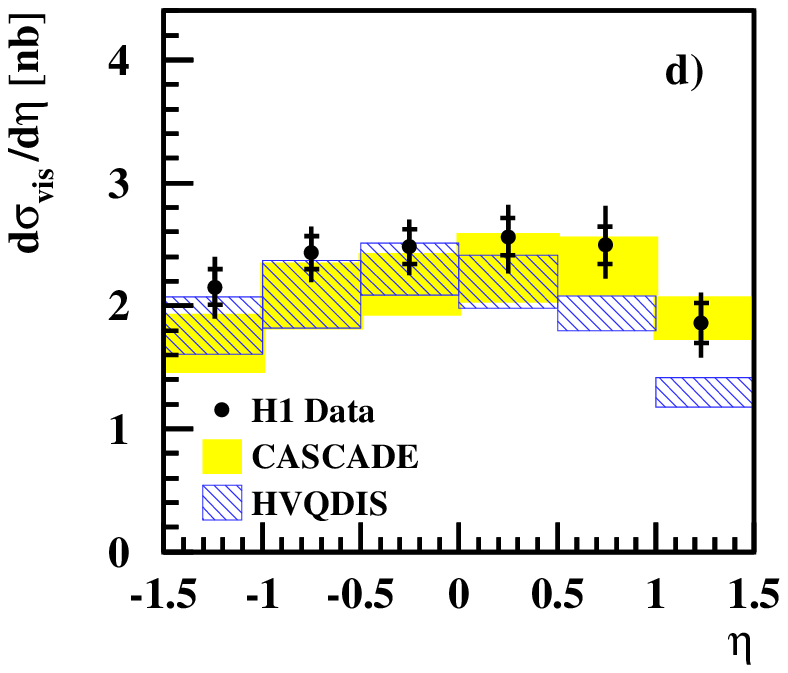,width=0.45\linewidth,bbllx=0,bblly=0,bburx=115,bbury=219}}
\end{minipage}
\vspace*{-6mm}
\break\hspace*{-7mm}
\begin{minipage}{8.10cm}
\mbox{\epsfig{file=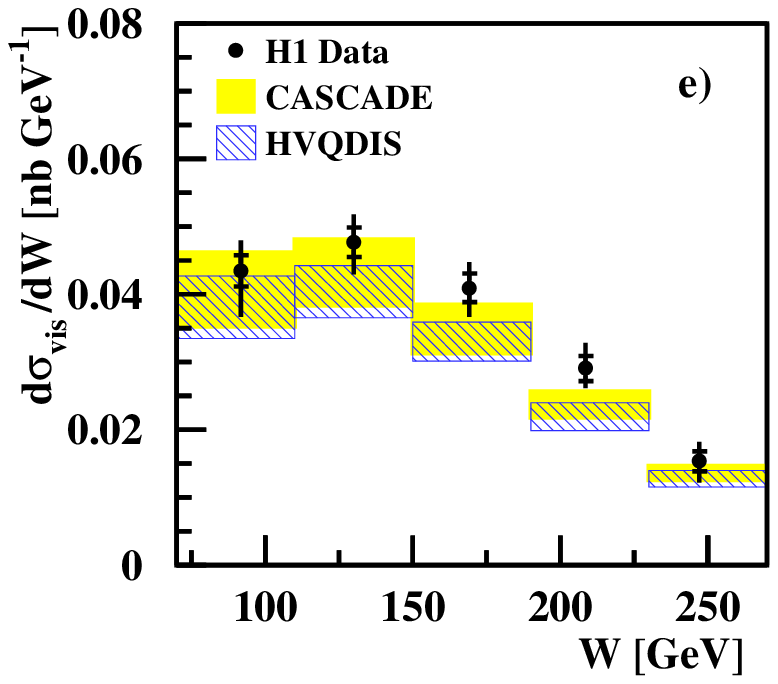,width=0.45\linewidth,bbllx=0,bblly=0,bburx=115,bbury=219}}
\end{minipage}
\begin{minipage}{8.10cm}
\mbox{\epsfig{file=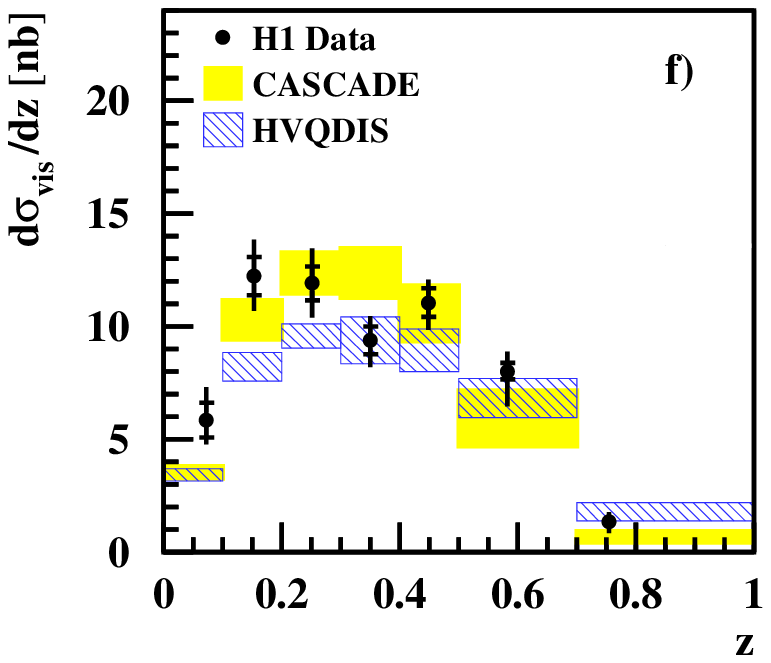,width=0.45\linewidth,bbllx=0,bblly=0,bburx=115,bbury=219}}
\end{minipage}
\vspace*{-9mm}
\end{center}
\vspace*{-1mm} \caption{\label{fig3a}{
Differential cross sections for inclusive $D^{*\pm}$ meson production
as a function of $Q^2$, $x$, $W$, $p_{\rm T}$, $\eta$ and $z$. The inner error bars indicate the statistical errors, and the outer error bars show the statistical and systematic uncertainties added in quadrature. The bands for the expectations of HVQDIS and CASCADE are obtained using the parameter variations as described in secion~2. Figures a), b) and c) also present the ratio 
$R=\sigma_{\rm theory} /\sigma_{\text{data}}$ for the predictions as bands, by taking into account their theoretical uncertainties. The inner error bars of the data points at $R=1$ display the relative statistical errors, and the outer error bars show the
relative statistical and systematic uncertainties added in quadrature. }}
\end{figure}

In figure~\ref{fig3a} the single differential cross sections for 
inclusive $D^{*\pm}$ production in the visible region are shown\footnote{The bin averaged cross section is shown at the position of the centre-of-gravity of the cross section in that bin as calculated by RAPGAP.} as a function of the event variables $Q^2$, $x$ and $W$ and of the $D^{*\pm}$ observables $p_{\rm T}$ and $\eta$ and of the
inelasticity $z$. The latter is defined as 
$z={P\cdot p}/{P\cdot q}={(E-p_z)_{D^*}}/{2yE_e}$, where $P$, $q$ and
$p$ denote the four-momenta of the incoming proton, the exchanged photon and the observed $D^{*\pm}$ meson, respectively. This quantity is a measure of the fraction of photon energy transferred to the
$D^{*\pm}$ meson in the proton rest frame and it is sensitive to both the production mechanism and the $c \ra D^{*\pm}$ fragmentation function. The measured cross sections shown in 
figure~\ref{fig3a} are listed in tables~\ref{txsec1} and~\ref{txsec2} and are in good agreement with previous measurements from 
H1~\cite{DESY01100}. Double differential cross sections as functions of $Q^2$ and $x$ are listed in table~\ref{txsec22}.

Figure~\ref{fig3a} includes the expectations from the HVQDIS and the CASCADE programs. The ratio $R$ of the theoretical to the measured cross section is also shown for selected distributions. The steep fall of the cross section as a function of $Q^2$ and $x$ is described by both HVQDIS and CASCADE. There is reasonable agreement between HVQDIS and the data for the different single differential cross sections with the exception of the medium values of $p_{\rm T}$ and the region
$\eta>0$, where the measured $D^{*\pm}$ meson production cross
section is larger than predicted. An excess is also observed at small $z$, a region correlated with the forward direction ($\eta>0$). Already in \cite{DESY01100} indications of an excess observed in the data at large pseudorapidities ($0.5<\eta<1.5$) and small $z$ with respect to the HVQDIS expectation had been reported. The data presented here confirm this excess with better statistical precision
and across the whole range of $Q^2$, as shown in a detailed study of correlations among the observables in $D^{*\pm}$ meson production \cite{sebthesis}. The predictions from the CASCADE program are found to generally agree better with the data than those from HVQDIS. An excess of the data over collinear NLO predictions at $\eta>0$ and small $z$ has been also observed in the photoproduction of $D^{*\pm}$ mesons\cite{gero}.

\begin{figure}
\begin{center}
{\Large {\bf H1 ${\;\; \mathbf{ep \ra eD^{*\pm}X}}$}}
\vspace*{1mm}
\break\hspace*{-7mm}
\begin{minipage}{8.10cm}
\mbox{\epsfig{file=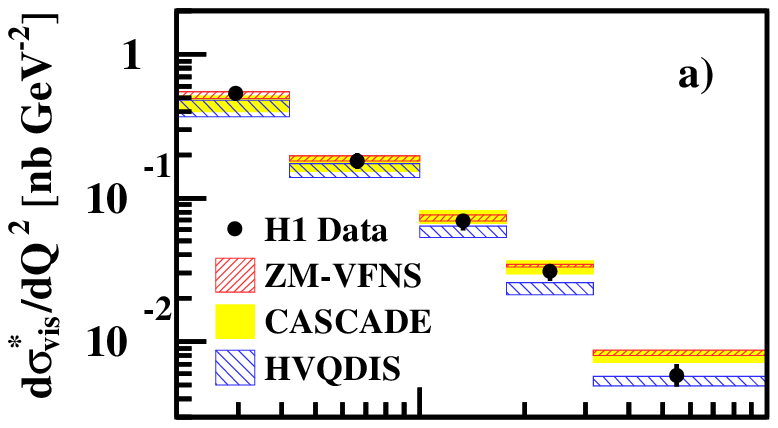,width=0.45\linewidth,bbllx=0,bblly=57,bburx=115,bbury=180}}\break
\mbox{\epsfig{file=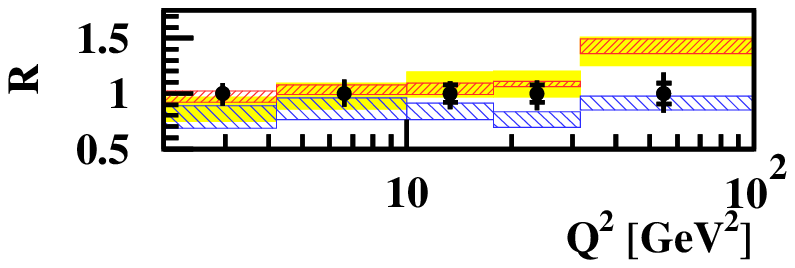,width=0.45\linewidth,bbllx=0,bblly=0,bburx=115,bbury=96}}
\end{minipage}
\begin{minipage}{8.10cm}
\mbox{\epsfig{file=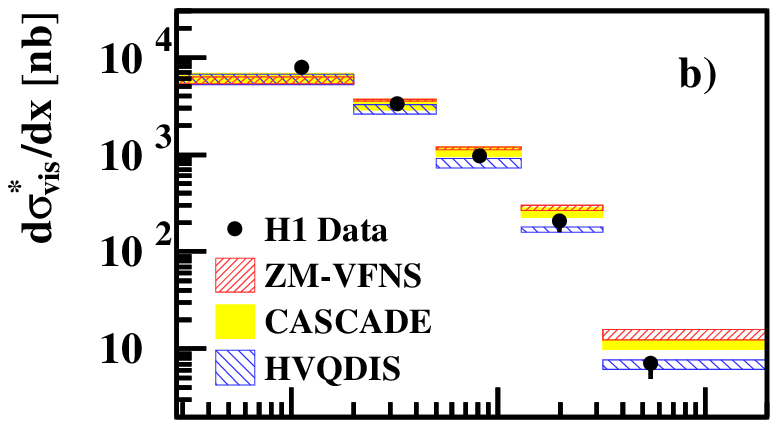,width=0.45\linewidth,bbllx=0,bblly=57,bburx=115,bbury=180}}\break
\mbox{\epsfig{file=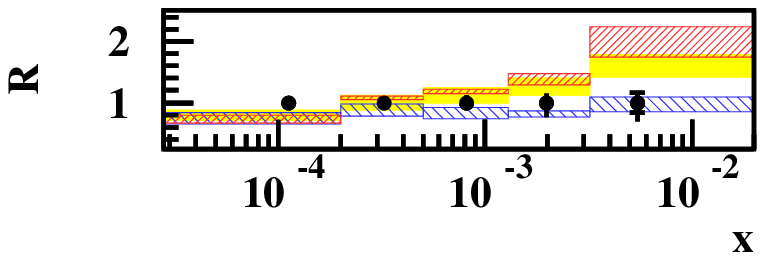,width=0.45\linewidth,bbllx=0,bblly=0,bburx=115,bbury=96}}
\end{minipage}
\vspace*{-5mm}
\break\hspace*{-7mm}
\begin{minipage}{8.10cm}
\mbox{\epsfig{file=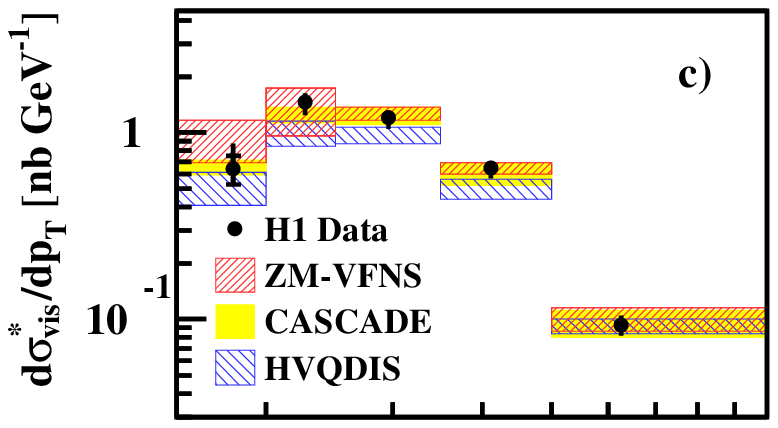,width=0.45\linewidth,bbllx=0,bblly=57,bburx=115,bbury=180}}\break
\mbox{\epsfig{file=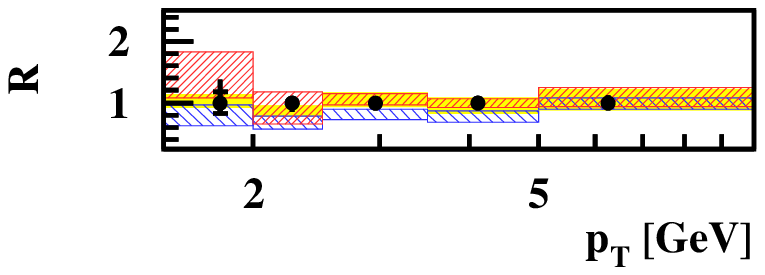,width=0.45\linewidth,bbllx=0,bblly=0,bburx=115,bbury=96}}
\end{minipage}
\begin{minipage}{8.10cm}
\mbox{\epsfig{file=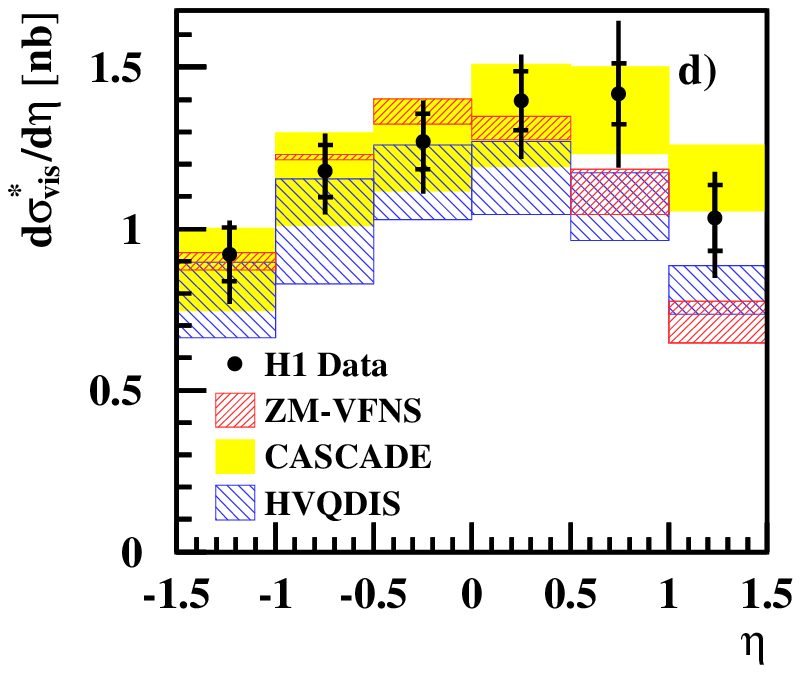,width=0.45\linewidth,bbllx=0,bblly=0,bburx=115,bbury=219}}
\end{minipage}
\vspace*{-10mm}
\end{center}
\caption{\label{fig3c}{
Differential cross sections for inclusive 
$D^{*\pm}$ meson production with, compared to figure~\ref{fig3a}, the additional requirement $p_{\rm T}^*>2.0$ GeV for the $D^{*\pm}$ meson in the $\gamma p$ center-of-mass frame as a function of $Q^2$, $x$, $p_{\rm T}$ and $\eta$. The inner error bars indicate the statistical errors, and the outer error bars show the statistical and systematic uncertainties added in quadrature. The bands for the expectations of ZM-VFNS, a ``massless'' QCD calculation ~\cite{massless,zmfrag} and of HVQDIS and CASCADE are obtained using the parameter variations as described in section~2. The ratio R is described in the caption of figure~\ref{fig3a}. }}
\end{figure}

In figure~\ref{fig3c} the inclusive $D^{*\pm}$ production cross section with the additional condition on the $D^{*\pm}$ meson
$p_{\rm T}^* > 2.0$ GeV, in order to be able to compare with the 
ZM-VFNS predictions, is shown. The predictions of CASCADE and particularly the one of the ZM-VFNS approach are not able to describe the data at large $x$ and large $Q^2$, while HVQDIS is consistent with the data within errors. The expectation of the ZM-VFNS calculation is very close to the one from CASCADE for $Q^2$ and $x$, however for 
the forward direction in $\eta$ it is lower than CASCADE and similar to the expectation from HVQDIS. The measured cross sections are listed in table~\ref{txsec3}.

\section{Production cross sections for $\mathbf D^{*\pm}$ Mesons with Dijets}

The production cross section of $D^{*\pm}$ mesons with dijets in the
kinematic region $2 \le Q^2 \le 100\ \mbox{GeV}^2$, 
$0.05 \le y \le 0.7$, in the visible $D^{*\pm}$ range 
$1.5 \le p_{\rm T} \le 15\ \mbox{GeV}$ and $|\eta| \le 1.5$, and with jets having Breit frame transverse energies 
$E_{\rm T}^{\rm jet 1(2)} \ge 4(3)\ \mbox{GeV}$,
and laboratory pseudo\-rapidities 
$-1 \le \eta_{\rm lab}^{\rm jet 1,2} \le 2.5$, is found to be
\begin{linenomath}\begin{equation}\label{eqsigmajj}
\sigma_{\rm vis}(e^+p \ra e^+D^{*\pm}\text{jj} \ X)= 1.60\pm
0.12\, (\rm{stat.})\, ^{+0.14} _{-0.18} \, (\rm{syst.})\ \rm{nb}.
\end{equation}\end{linenomath}
The predictions from HVQDIS and CASCADE are listed in 
table~\ref{tabinclpred}. As for the inclusive $D^{*\pm}$ meson analysis the nominal HVQDIS value is lower ($16\%$) than the data, but there is agreement within errors. The prediction by CASCADE agrees well with the data. The uncertainty of the two predictions due to fragmentation is much reduced compared to the inclusive $D^{*\pm}$ meson case.
For CASCADE, also the scale uncertainty is reduced while for HVQDIS it remains the dominant error contribution which is found to be even 
larger than for the inclusive production of $D^{*\pm}$ mesons.
Use of the MRST2004F3NLO instead of the CTEQ5F3 parton densities of the proton for HVQDIS results in a $1\%$ increase of the cross section. If for the $k_{\rm T}$-unintegrated gluon distribution of the proton in CASCADE the J2003 set-1~\cite{gluonj2003} parametrisation is used, instead of set-A0, the cross section increases by $1\%$.

The systematic uncertainty is reduced significantly in the ratio of the production cross section for $D^{*\pm}$ mesons with dijets to the one for inclusive $D^{*\pm}$ mesons
\begin{linenomath}\begin{equation}\label{eqratio}
\frac{\sigma_{\rm vis}(e^+p \ra e^+D^{*\pm}\text{jj} \ X)}
{\sigma_{\rm vis}(e^+p \ra e^+D^{*\pm} \ X)}= 0.228\pm 0.014\,
(\rm{stat.}) \pm 0.011 \, (\rm{syst.}).
\end{equation}\end{linenomath}
As shown in table~\ref{tabinclpred} the expectation from the HVQDIS program agrees well with the measured ratio while CASCADE predicts a somewhat larger value.

\begin{figure}
\begin{center}
{\Large {\bf H1 ${\;\; \mathbf{ep \ra eD^{*\pm}jjX}}$}}
\vspace*{1mm}
\break\hspace*{-7mm}
\begin{minipage}{8.10cm}
\mbox{\epsfig{file=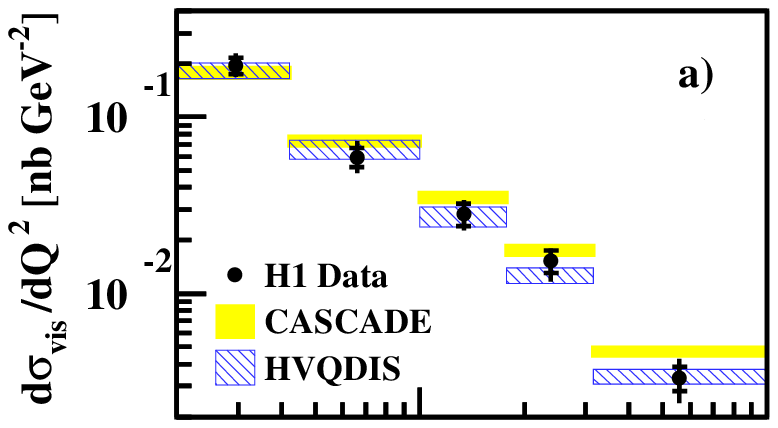,width=0.45\linewidth,bbllx=0,bblly=57,bburx=115,bbury=180}}\break
\mbox{\epsfig{file=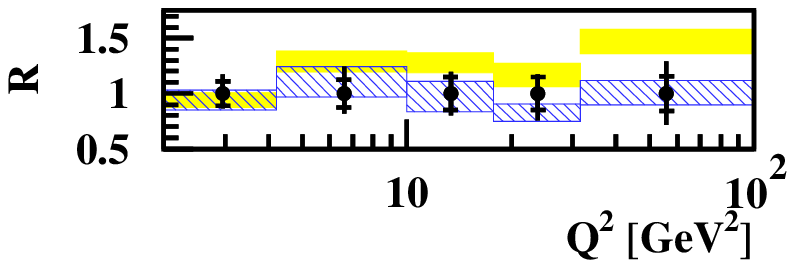,width=0.45\linewidth,bbllx=0,bblly=0,bburx=115,bbury=96}}
\end{minipage}
\begin{minipage}{8.10cm}
\mbox{\epsfig{file=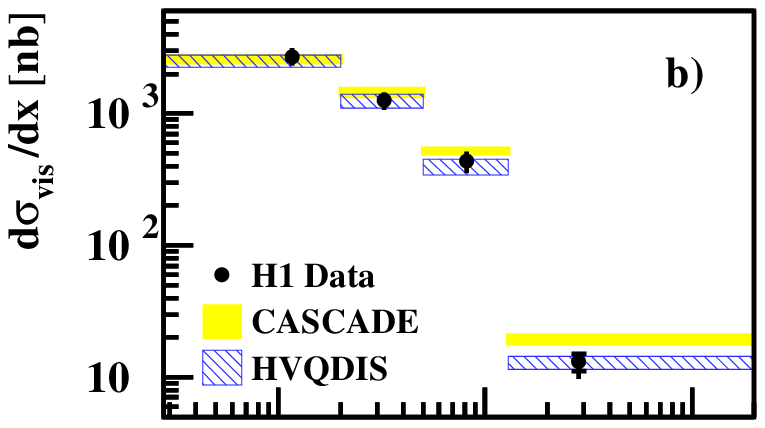,width=0.45\linewidth,bbllx=0,bblly=57,bburx=115,bbury=180}}\break
\mbox{\epsfig{file=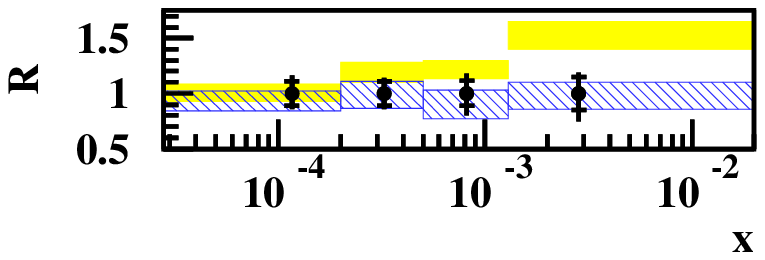,width=0.45\linewidth,bbllx=0,bblly=0,bburx=115,bbury=96}}
\end{minipage}
\vspace*{-5mm}
\break\hspace*{-7mm}
\begin{minipage}{8.10cm}
\mbox{\epsfig{file=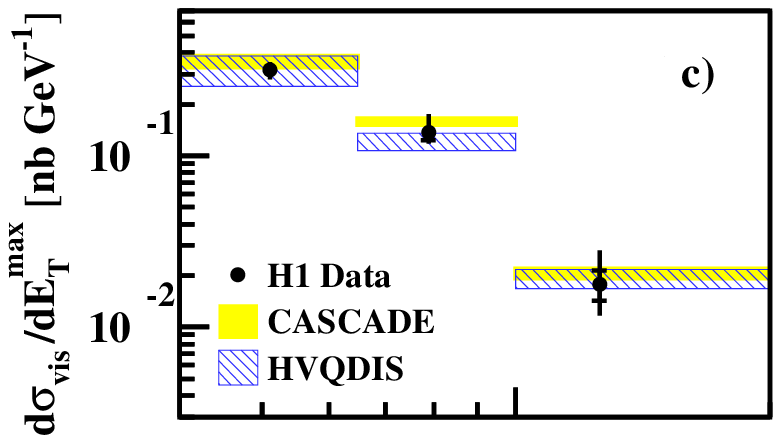,width=0.45\linewidth,bbllx=0,bblly=57,bburx=115,bbury=180}}\break
\mbox{\epsfig{file=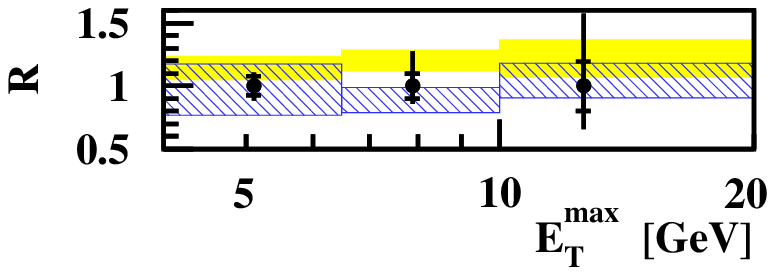,width=0.45\linewidth,bbllx=0,bblly=0,bburx=115,bbury=96}}
\end{minipage}
\begin{minipage}{8.10cm}
\mbox{\epsfig{file=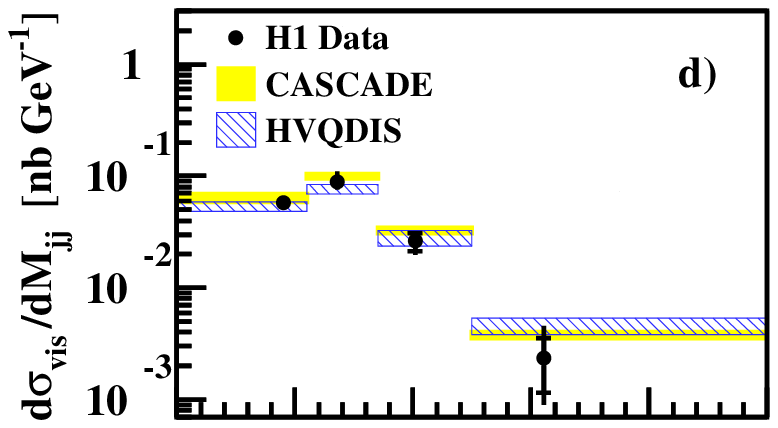,width=0.45\linewidth,bbllx=0,bblly=57,bburx=115,bbury=180}}\break
\mbox{\epsfig{file=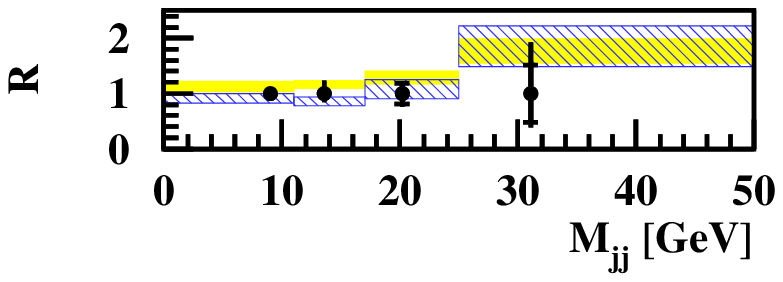,width=0.45\linewidth,bbllx=0,bblly=0,bburx=115,bbury=96}}
\end{minipage}
\end{center}
\vspace*{-10mm}
\caption{\label{jetxsec}{
Differential cross sections for the production of $D^{*\pm}$ mesons with dijets as a function of $Q^2$, $x$, $E_{\rm T}^{\text{max}}$ in the Breit frame and $M_{\rm jj}$. The inner error bars indicate the statistical errors, and the outer error bars show the statistical and systematic uncertainties added in quadrature. The bands for the expectations of HVQDIS and CASCADE are obtained using the parameter variations as described in section~2. The ratio R is described in the caption of figure~\ref{fig3a}. }}
\end{figure}

In figure~\ref{jetxsec} the differential cross sections for $D^{*\pm}$ mesons with dijets are presented as functions of the event variables $Q^2$ and $x$, and of the jet variables
$E_{\rm T}^{\rm max} = E_{\rm T}^{\rm jet 1}$ which is the maximum transverse jet energy in the Breit frame and the invariant mass 
$M_{\rm jj}$ of the dijet system. The data are compared with the expectations from HVQDIS and CASCADE indicated by bands, which indicate the total theoretical uncertainty as described in section~2. 
The uncertainty on $m_c$ is responsible for the large error of the HVQDIS prediction for low values of 
$E_{\rm T}^{\rm max}$. The quality of the description is more obvious in the ratio of the predicted to the measured cross sections, which is also shown in figure~\ref{jetxsec}. Both HVQDIS and CASCADE describe the steep fall of the cross section as $Q^2$ and $x$ become large, though CASCADE systematically overestimates the cross section at high values of $Q^2$ and $x$. The distributions of $E_{\rm T}^{\rm max}$ and $M_{\rm jj}$ are well described by both HVQDIS and CASCADE.

\begin{figure}
\begin{center}
{\Large {\bf H1 ${\;\; \mathbf{ep \ra eD^{*\pm}jjX}}$}}
\vspace*{-5mm}
\break\hspace*{-12mm}
\begin{minipage}{8.10cm}
\mbox{\epsfig{file=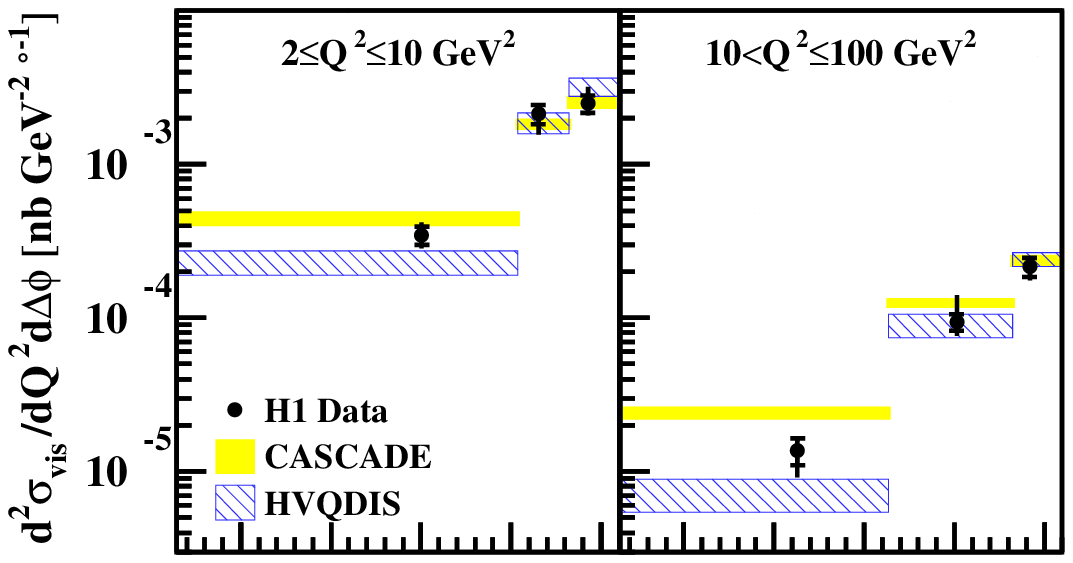,width=0.45\linewidth,bbllx=40,bblly=58,bburx=150,bbury=234}}\break
\mbox{\epsfig{file=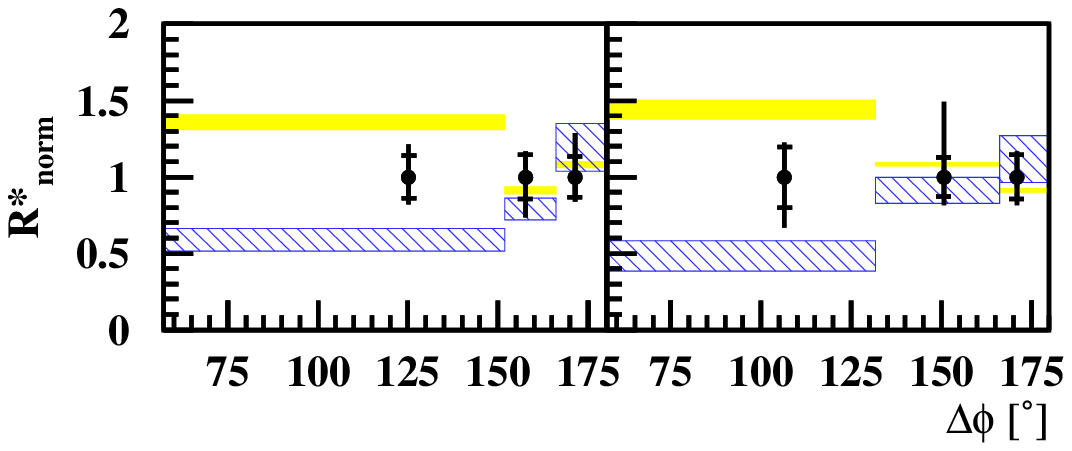,width=0.45\linewidth,bbllx=40,bblly=0,bburx=150,bbury=145.5}}
\end{minipage}
\end{center}
\vspace*{-10mm}
\caption{\label{jetxsecdphi}{
Double differential cross sections for the production of $D^{*\pm}$ mesons with dijets as a function of $\Delta\phi$ in the Breit frame for two regions in $Q^2$. The inner error bars indicate the statistical errors, and the outer error bars show the statistical and systematic uncertainties added in quadrature. The bands for the expectations of HVQDIS and CASCADE are obtained using the parameter variations as described in section~2. Also shown is the ratio $R_{\text{norm}}^*$, for which the cross section in the last two bins is used for normalisation (for details see section~7). }}
\end{figure}

The absolute difference in azimuthal angle in the Breit frame,
$\Delta\phi=|\phi_{\text{jet 1}}-\phi_{\text{jet 2}}|$, is shown in figure~\ref{jetxsecdphi} as a double differential cross section for two ranges of $Q^2$. In the LO photon-gluon fusion process 
($\gamma g \ra c \bar{c}$) the two jets are expected to be 
back-to-back, i.e. $\Delta\phi = 180^\circ$. The contributions at 
$\Delta\phi < 180^\circ$ arise primarily from hard gluon emissions and fragmentation effects. HVQDIS allows radiation of one hard gluon in NLO, while CASCADE includes the radiation of one or more gluons in the parton showering process and it contains transverse momentum $k_{\rm T}$ effects of the exchanged gluons. The measured cross sections for large $\Delta\phi$ (bin two and three) are well described by both theoretical approaches as expected. In order to be sensitive to higher order or $k_{\rm T}$ effects at smaller 
$\Delta\phi$, and because theoretical uncertainties are reduced, the following ratio is defined
\begin{linenomath}\begin{equation}\label{rnormstar}
R_{\text{norm}}^* = \frac{
\frac{ {\rm d}^2\sigma_{\rm vis}^{\rm theory}}{{\rm d}Q^2{\rm d}\Delta\phi} } {
\int\limits_{{\Delta\phi{\rm (bin \,} 2+3{\rm )}} }
\frac{ {\rm d}^2\sigma_{\rm vis}^{\rm theory}}{{\rm d}Q^2{\rm d}\Delta\phi} } {\Bigg /} \frac{
\frac{ {\rm d}^2\sigma_{\rm vis}^{\rm data}}{{\rm d}Q^2{\rm d}\Delta\phi} } {
\int\limits_{{\Delta\phi{\rm (bin \,} 2+3{\rm )}} } 
\frac{ {\rm d}^2\sigma_{\rm vis}^{\rm data}}{{\rm d}Q^2{\rm d}\Delta\phi} } \;\; .
\end{equation}\end{linenomath}

This double ratio of theory over data is also shown in figure~\ref{jetxsecdphi} as well as the data points to indicate the experimental errors.
At the lowest values of $\Delta\phi$ (bin $1$), CASCADE is slightly above the data, indicating that the $k_{\rm T}$-distribution in the unintegrated gluon density in CASCADE is too broad. HVQDIS, on the other hand, underestimates the cross section at the lowest 
$\Delta\phi$, both at lower and at higher $Q^2$, indicating that in this approach effects beyond NLO are needed to match the data. Similar conclusions were obtained in photoproduction from measurements of the azimuthal correlations of a $D^{*\pm}$ meson and a jet not associated
to the $D^{*\pm}$ meson \cite{gero} and of dijets in events with a 
$D^{*\pm}$ meson \cite{zeusdphi}. The measured cross sections are listed in table~\ref{txsec4}.

\begin{figure}
\begin{center}
{\Large {\bf H1 ${\;\; \mathbf{ep \ra eD^{*\pm}jjX}}$}}
\vspace*{-5mm}
\break\hspace*{-12mm}
\begin{minipage}{8.10cm}
\mbox{\epsfig{file=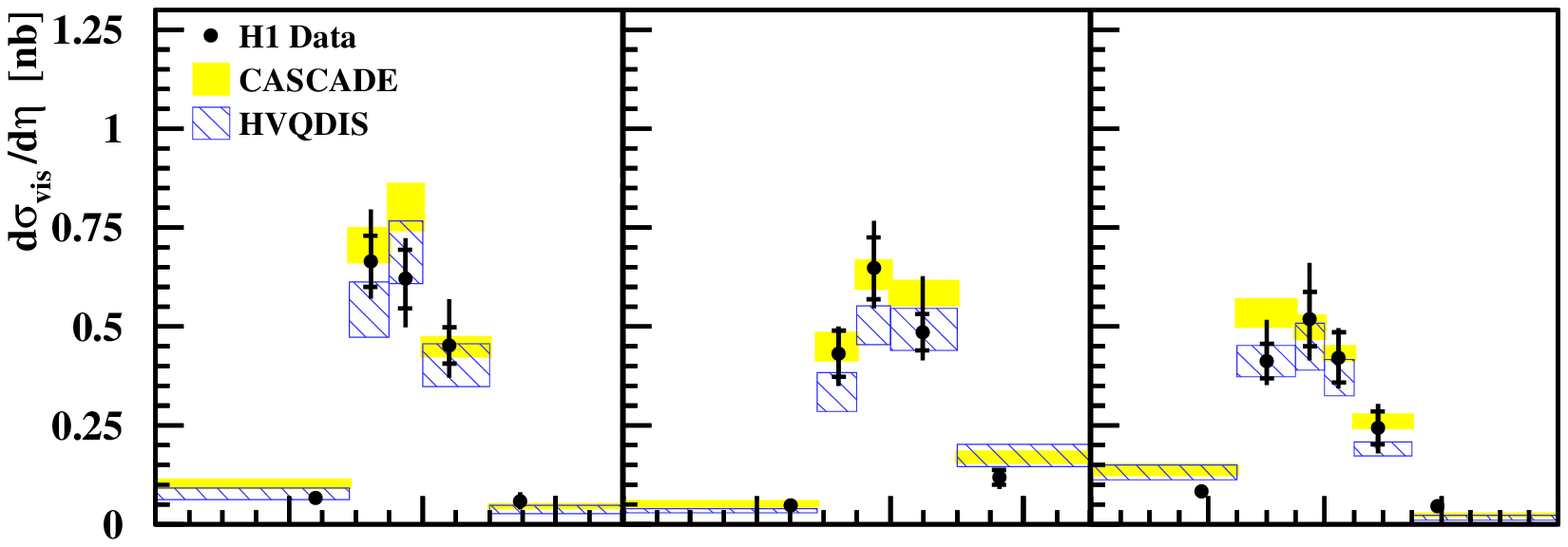,width=0.70\linewidth,bbllx=110,bblly=58,bburx=296,bbury=234}}\break
\mbox{\epsfig{file=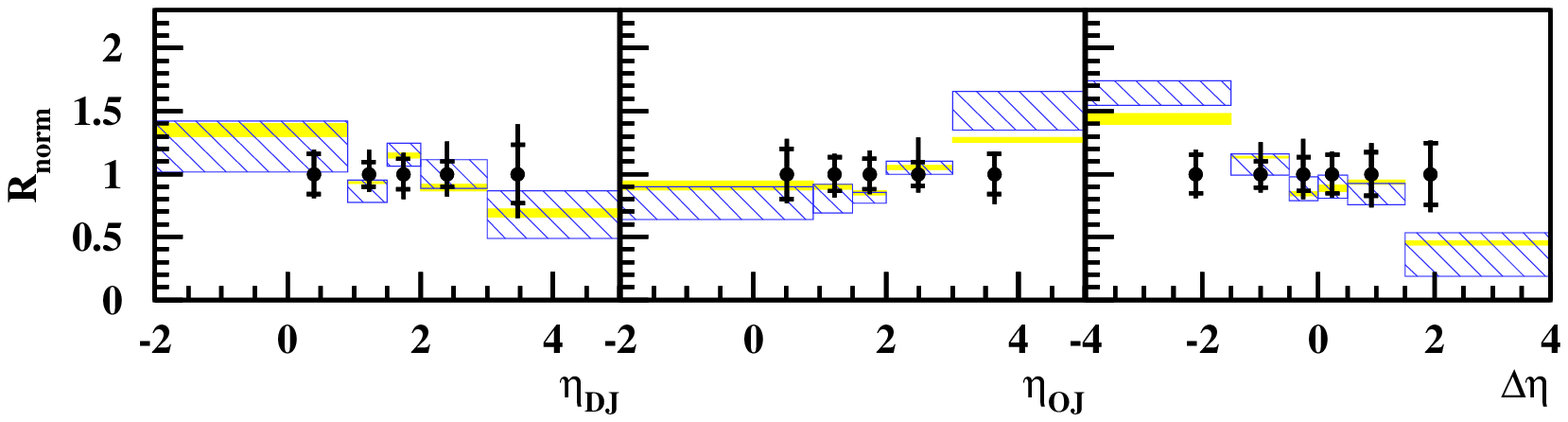,width=0.70\linewidth,bbllx=110,bblly=0,bburx=296,bbury=145.5}}
\end{minipage}
\vspace*{-10mm}
\end{center}
\caption{\label{djetxsec}{
Differential cross sections for the production of $D^{*}$-jet and other jet as a function of the pseudorapidity of the $D^*$-jet and the other jet (OJ) and of the difference in pseudorapidity 
$\Delta\eta=\eta_{\rm DJ} - \eta_{\rm OJ}$ of the two jets. The inner error bars indicate the statistical errors, and the outer error bars show the statistical and systematic uncertainties added in quadrature. The bands for the expectations of HVQDIS and CASCADE are obtained using the parameter variations as described in section~2. In addition, the ratio $R_{\text{norm}}$ is shown (for details see section~7). }}
\end{figure}

The jet which contains the $D^{*\pm}$ meson, $D^{*}$-jet (DJ)\footnote{It is typically one of the two leading jets. However this is not required in which case the only constraint on the $D^{*}$-jet is due to the $D^{*\pm}$ meson kinematic cuts.}, and the other jet (OJ) with the highest $E_{\rm T}^{\rm jet}$ not containing the $D^{*\pm}$ meson are further investigated. The OJ allows a larger region in rapidity to be accessed, the more forward direction, compared to the DJ. In 
figure~\ref{djetxsec} the cross section is shown as a function of the pseudorapidity of the $D^*$-jet, the other jet, and of the difference in pseudorapidity of the two jets, 
$\Delta\eta=\eta_{\rm DJ} - \eta_{\rm OJ}$, all measured in the Breit frame. The pseudorapidity distributions are reasonably reproduced by both HVQDIS and CASCADE.
While the region of small values of $|\Delta\eta|$, which might be expected to be particularly sensitive to low-$x$ dynamics~\cite{lowxref}, is well described, small discrepancies are observed for forward going other jets and for large $|\Delta\eta|$.
These are more clearly seen in the ratio $R_{\rm norm}$\footnote{Here all bins are used for the normalisation in contrast to $R^{*}_\text{norm}$.} which has reduced theoretical uncertainty. The measured cross sections are listed in table~\ref{txsec5}.

To further improve the understanding of the charm production mechanism in DIS, the observables $x_{\gamma}^{\rm obs}$ and 
$x_{\rm g}^{\rm obs}$ are investigated. At LO they give the observed
fraction of the photon momentum carried by the parton involved in the hard subprocess and the observed fraction of the proton momentum carried by the gluon, respectively. The determination of both quantities involves the partons emerging from the hard subprocess, which are approximated by the $D^*$-jet and the other jet.

\begin{figure}
\begin{center}
{\Large {\bf H1 ${\;\; \mathbf{ep \ra eD^{*\pm}jjX}}$}}
\vspace*{-5mm}
\break
\mbox{\epsfig{file=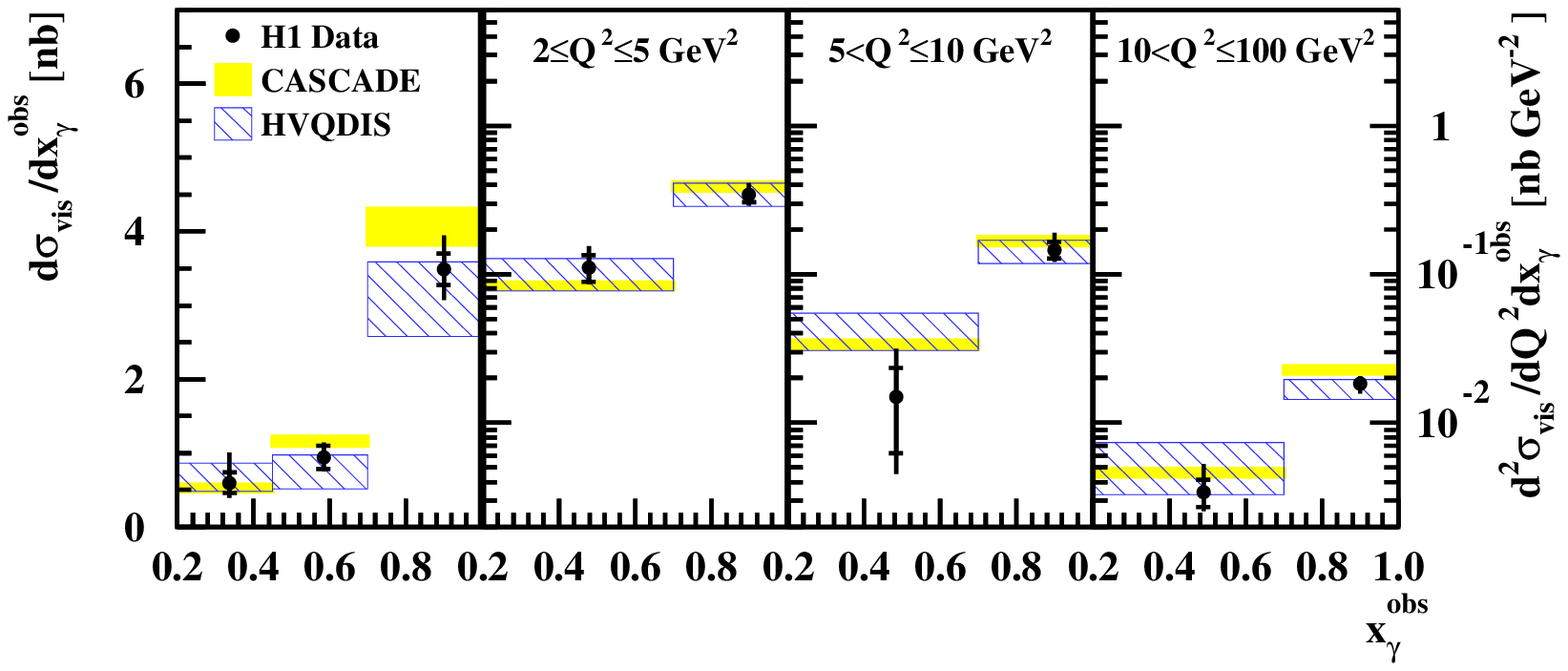,clip=,width=15.6cm,bbllx=25,bblly=0,bburx=484,bbury=234}}\break
\mbox{\epsfig{file=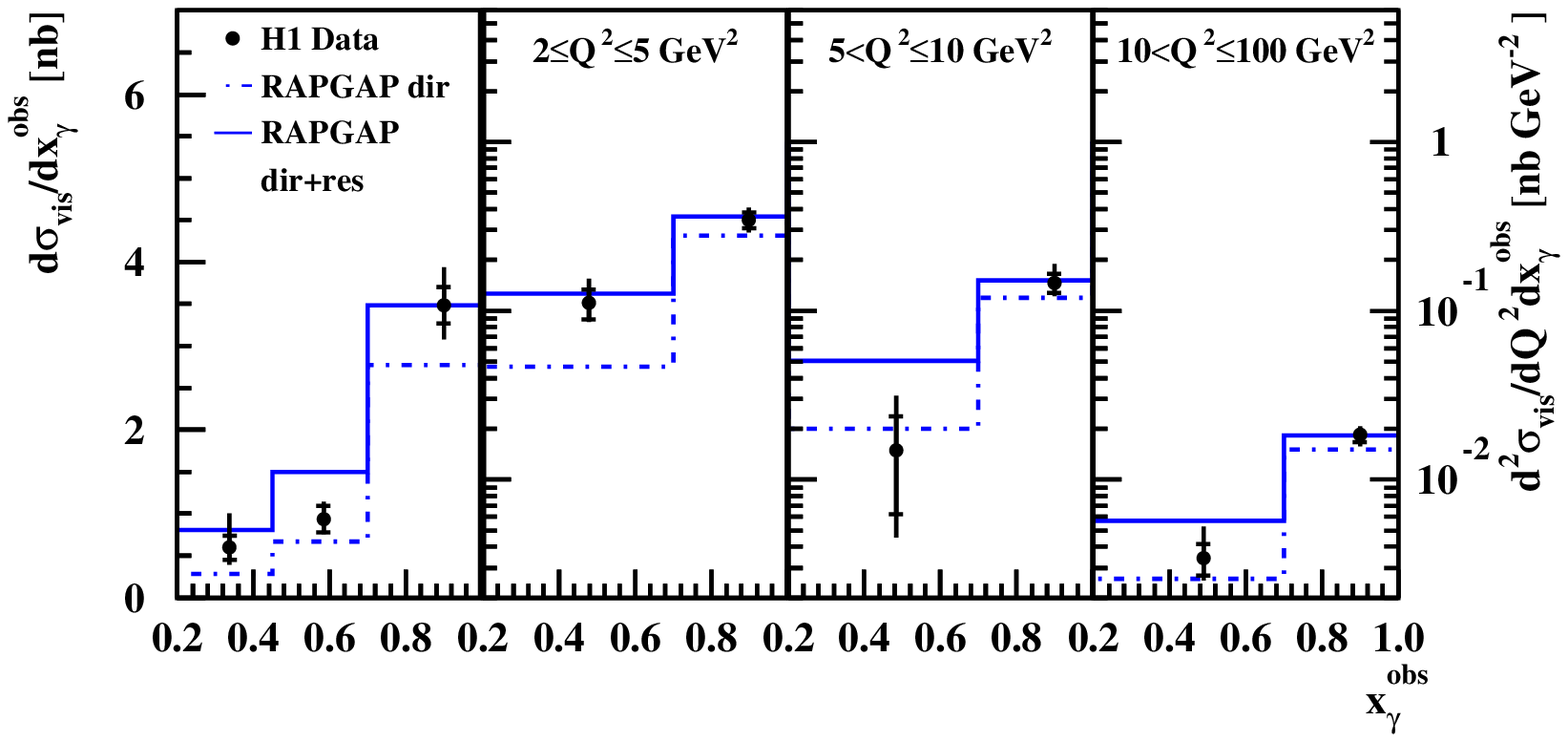,clip=,width=15.6cm,bbllx=25,bblly=0,bburx=484,bbury=234}}
\end{center}
\vspace*{-10mm}
\caption{\label{fig_diffxgamma}{
Differential cross sections for the production of $D^{*}$-jet and other jet as a function of $x_{\gamma}$ and double differentially in three $Q^2$ regions. The inner error bars indicate the statistical errors, and the outer error bars show the statistical and systematic uncertainties added in quadrature. The bands for the expectations of HVQDIS and CASCADE are obtained using the parameter variations as described in section~2. For 
$x_{\gamma}^{\rm obs}$ the uncertainty of the hadronization correction is the dominant contribution to the total theoretical uncertainty. In the lower plot, the RAPGAP prediction, indicating the direct and the sum of direct and resolved contributions, is shown. }}
\end{figure}

The observable $x_{\gamma}^{\rm obs}$ is defined as
\begin{linenomath}\begin{equation}\label{xgamma}
x_{\gamma}^{\rm obs} = \frac{(E^*-p^*_z)_{\rm DJ} + 
(E^*-p^*_z)_{\rm OJ}}{(E^*-p^*_z)_{\rm had}},
\end{equation}\end{linenomath}
where $E^*$ and $p^*_z$ are measured in the $\gamma p$ center-of-mass frame. In the numerator $(E^*-p^*_z)$ is summed over all particles belonging to the two jets and in the denominator 
$(E^*-p^*_z)_{\rm had}$ is the sum over all hadronic final state objects. In figure~\ref{fig_diffxgamma} the single differential cross section for the production of dijets with a 
$D^{*\pm}$ meson is shown as a function of $x_{\gamma}^{\rm obs}$ and double differentially in three bins of $Q^2$. The distribution of $x_{\gamma}^{\rm obs}$ peaks close to 1 as expected from direct processes, but has significant contributions at lower values. The HVQDIS predictions are in reasonable agreement with the measured cross section as a function of $x_{\gamma}^{\rm obs}$, and they describe the $Q^2$ dependence of $x_{\gamma}^{\rm obs}$, indicating that there is no need for an additional resolved photon contribution beyond what is already included at NLO. CASCADE also provides a reasonable description.
The expectation by RAPGAP with direct and resolved contributions is
similar to the HVQDIS prediction. For $Q^2>5$~GeV$^2$ the data can be described by multiplying the RAPGAP direct contribution by a constant factor, independent of $x_{\gamma}^{\rm obs}$. However, for $Q^2<5$ GeV$^2$ the data indicate that a constant factor would not be sufficient, and only the addition of a resolved photon component 
leads to a good description in LO models based on collinear factorisation.

\begin{figure}
\begin{center}
{\Large {\bf H1 ${\;\; \mathbf{ep \ra eD^{*\pm}jjX}}$}}
\vspace*{-3mm}
\break
\begin{minipage}{8.10cm} \hspace*{0.5mm}
\mbox{\epsfig{file=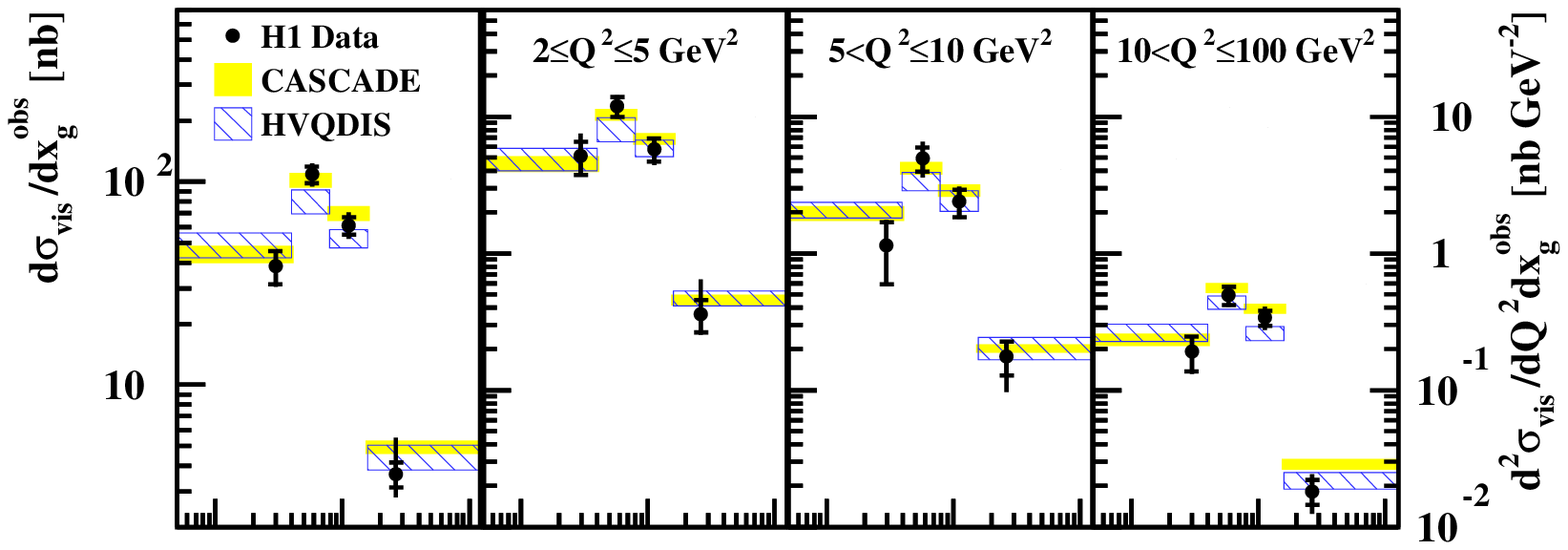,clip=,width=18.6cm,bbllx=140,bblly=46,bburx=680,bbury=234}}\break \hspace*{0.5mm}
\mbox{\epsfig{file=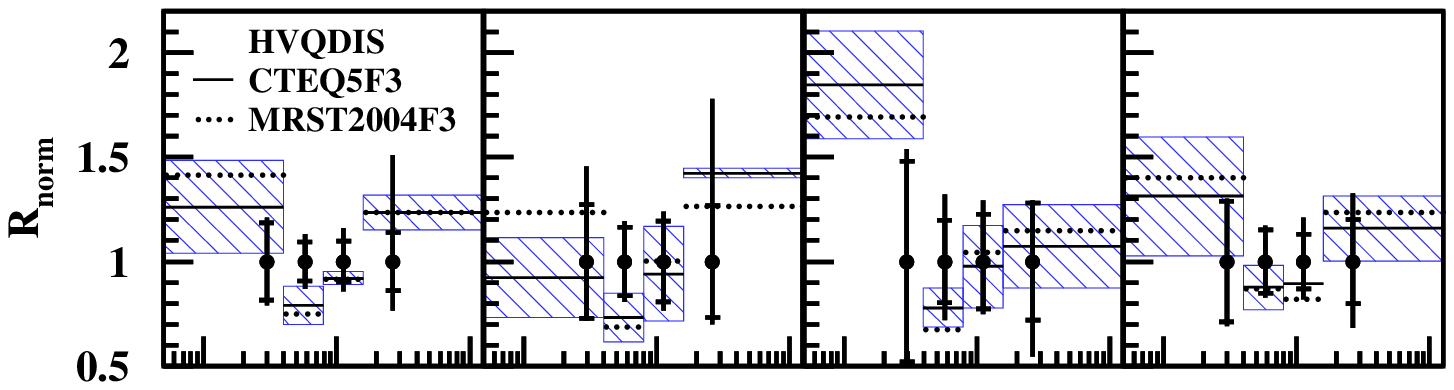,clip=,width=18.6cm,bbllx=140,bblly=46,bburx=680,bbury=147.5}}\break \hspace*{0.5mm}
\mbox{\epsfig{file=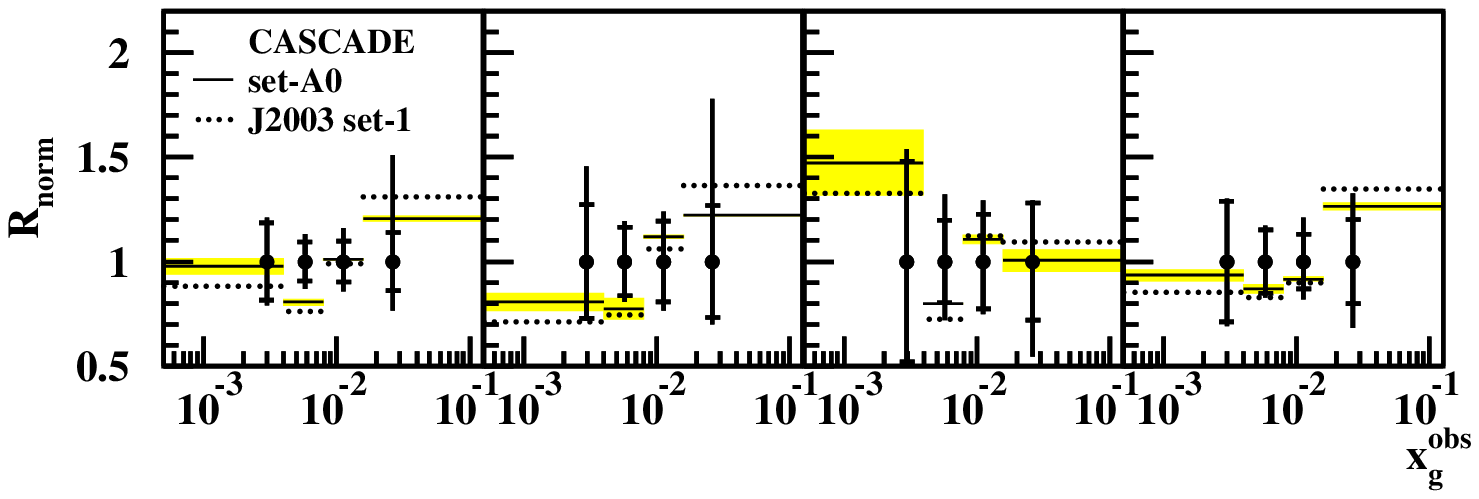,clip=,width=18.6cm,bbllx=140,bblly=0,bburx=680,bbury=147.5}}
\end{minipage}
\end{center}
\vspace*{-10mm}
\caption{\label{fig_diffxgluon}{
Differential cross sections for the production of $D^{*}$-jet and other jet as a function of $x_{\rm g}^{\rm obs}$ and double differentially in three $Q^2$ regions. The inner error bars indicate the statistical errors, and the outer error bars show the statistical and systematic uncertainties added in quadrature. The bands for the expectations of HVQDIS and CASCADE are obtained using the parameter variations as described in section~2. 
The ratio $R_{\text{norm}}$ is also shown, separately for HVQDIS and CASCADE. For the HVQDIS (CASCADE) band the CTEQ5F3 (set-A0) PDF is used, the central predictions being given by the full line. The central values using the PDF MRST2004F3NLO (J2003 set-1) are indicated as dotted line. }}
\end{figure}

The observable $x_{\rm g}^{\rm obs}$ is defined as
\begin{linenomath}\begin{equation}\label{xgluon}
x_{\rm g}^{\rm obs} = \frac{E^*_{\rm T,DJ} e^{\eta^*_{\rm DJ}} +
E^*_{\rm T,OJ} e^{\eta^*_{\rm OJ}}}{2E_p^*}.
\end{equation}\end{linenomath}
The single differential cross section for $D^{*\pm}$ meson and dijet production is displayed in figure~\ref{fig_diffxgluon} as a function of $x_{\rm g}^{\rm obs}$ and double differentially in three regions of $Q^2$. The ratio $R_{\text{norm}}$ has reduced theoretical uncertainty and is also shown in figure~\ref{fig_diffxgluon}. Both HVQDIS and CASCADE with the parameter settings and the parton density functions listed in 
table~\ref{tabsettings} describe the $Q^2$ dependence of 
$x_{\rm g}^{\rm obs}$. The sensitivity to recent parton density parametrisations has been investigated by comparing with the predictions of HVQDIS using the MRST2004F3NLO parametrisation and the parametrisations set-B~\cite{hancassets} and J2003 set-1 for the
unintegrated gluon density in CASCADE. The differences for the various PDFs are small compared to the large uncertainties of the data. In figure~\ref{fig_diffxgluon} the predictions for $R_{\text{norm}}$ using for example MRST2004F3NLO and J2003 set-1 are compared to the default expectations using CTEQ5F3 and set-A0 with HVQDIS and CASCADE, respectively. The measured cross sections in bins of 
$x_{\rm g}^{\rm obs}$ and $x_{\gamma}^{\rm obs}$ are listed in 
tables~\ref{txsec6} and~\ref{txsec7}.

\section{Summary}

Measurements of the total and differential cross sections for inclusive $D^{*\pm}$ production in deep-inelastic $ep$ scattering are presented. In addition, cross sections are measured for dijets produced in events with a $D^{*\pm}$ meson. The general features of the data are described by the QCD predictions in the FFNS as implemented in HVQDIS and CASCADE, which are based on either collinear factorisation and DGLAP evolution or $k_{\rm T}$-factorisation and CCFM evolution, respectively.

Overall, CASCADE matches the inclusive $D^{*\pm}$ data somewhat better in normalisation and shape, and in particular in the positive pseudorapidity region. Furthermore the prediction of a calculation in the ZM-VFNS, where the charm quark is treated as massless is confronted with the data. It is also found to yield a satisfactory description.

Cross sections for the production of $D^{*\pm}$ mesons with dijets in DIS are presented as a function of various event and jet kinematic variables. Both HVQDIS and CASCADE give reasonable descriptions 
of the differential cross sections with HVQDIS providing a slightly better match to the data. The discrepancy observed between both calculations and data for azimuthal differences of the two leading jets $\Delta\phi$ below $150^{\circ}$ indicates that the 
$k_{\rm T}$-distribution in the unintegrated gluon density in
CASCADE is too broad, and that in the approach of HVQDIS 
effects beyond NLO are needed to match the data.

The $x_{\gamma}^{\rm obs}$ dependence of the cross section is described within the experimental and theoretical uncertainties by HVQDIS, indicating that there is no need for an additional resolved photon contribution beyond what is already included at NLO.
A reasonable description is obtained also by CASCADE. In addition, the $x_{\rm g}^{\rm obs}$ distribution is in agreement with the QCD predictions. This confirms that the input gluon distributions to the models, obtained from fits to inclusive data, provide a good representation of the charm data.

\section*{Acknowledgments}

We are grateful to the HERA machine group whose outstanding
efforts have made this experiment possible.
We thank
the engineers and technicians for their work in constructing and
maintaining the H1 detector, our funding agencies for
financial support, the
DESY technical staff for continual assistance
and the DESY directorate for support and for the
hospitality which they extend to the non DESY
members of the collaboration.

%
%

\clearpage
{\large
\begin{table}
 \begin{center}
\vspace*{-7mm}
{\footnotesize
\begin{tabular*}{10.0cm}{@{\extracolsep{0pt}}ll@{\extracolsep{\fill}}r@{\extracolsep{0pt}}l}
\noalign{\smallskip}&\begin{minipage}{2.7cm}$Q^2\,[\text{GeV}^2]$\hfill\end{minipage} & $\text{d}\sigma_\text{vis}(e^+p\rightarrow e^+D^{\ast\pm} X)/\text{d}Q^2\,[\text{nb}\,\text{GeV}^{-2}]$ & \cr\noalign{\smallskip}\hline\hline\noalign{\smallskip}
&$\left [2,\,4.22\right ]$ &  0.92\,$\pm$\,0.05$^{+0.07}_{-0.08}$ & \cr\noalign{\smallskip}
&$\left ]4.22,\,10\right ]$ &  0.358\,$\pm$\,0.017$^{+0.026}_{-0.029}$ & \cr\noalign{\smallskip}
&$\left ]10,\,17.8\right ]$ &  0.142\,$\pm$\,0.009$^{+0.012}_{-0.012}$ & \cr\noalign{\smallskip}
&$\left ]17.8,\,31.7\right ]$ &  0.065\,$\pm$\,0.004$^{+0.005}_{-0.006}$ & \cr\noalign{\smallskip}
&$\left ]31.7,\,100\right ]$ &  0.0124\,$\pm$\,0.0009$^{+0.0014}_{-0.0011}$ & \cr\noalign{\smallskip}
\hline\hline\noalign{\smallskip}
\end{tabular*}
}
{\footnotesize
\begin{tabular*}{10.0cm}{@{\extracolsep{0pt}}ll@{\extracolsep{\fill}}r@{\extracolsep{0pt}}l}
\noalign{\smallskip}&\begin{minipage}{2.7cm}$x$\hfill\end{minipage} & $\text{d}\sigma_\text{vis}(e^+p\rightarrow e^+D^{\ast\pm} X)/\text{d}x\,[\text{nb}]$ & \cr\noalign{\smallskip}\hline\hline\noalign{\smallskip}
&$\left [2.8 \cdot 10^{\text{-5}},\,0.0002\right ]$ &  11940\,$\pm$\,  587$^{+ 1079}_{- 1142}$ & \cr\noalign{\smallskip}
&$\left ]0.0002,\,0.0005\right ]$ &  6311\,$\pm$\, 311$^{+ 466}_{- 491}$ & \cr\noalign{\smallskip}
&$\left ]0.0005,\,0.0013\right ]$ &  2176\,$\pm$\, 106$^{+ 170}_{- 167}$ & \cr\noalign{\smallskip}
&$\left ]0.0013,\,0.0032\right ]$ &  498\,$\pm$\, 33$^{+ 36}_{- 54}$ & \cr\noalign{\smallskip}
&$\left ]0.0032,\,0.02\right ]$ &  18.4\,$\pm$\, 2.3$^{+ 1.6}_{- 2.4}$ & \cr\noalign{\smallskip}
\hline\hline\noalign{\smallskip}
\end{tabular*}
}
{\footnotesize
\begin{tabular*}{10.0cm}{@{\extracolsep{0pt}}ll@{\extracolsep{\fill}}r@{\extracolsep{0pt}}l}
\noalign{\smallskip}&\begin{minipage}{2.7cm}$W\,[\text{GeV}]$\hfill\end{minipage} & $\text{d}\sigma_\text{vis}(e^+p\rightarrow e^+D^{\ast\pm} X)/\text{d}W\,[\text{nb}\,\text{GeV}^{-1}]$ & \cr\noalign{\smallskip}\hline\hline\noalign{\smallskip}
&$\left [70,\,110\right ]$ &  0.043\,$\pm$\,0.002$^{+0.004}_{-0.006}$ & \cr\noalign{\smallskip}
&$\left ]110,\,150\right ]$ &  0.048\,$\pm$\,0.002$^{+0.004}_{-0.004}$ & \cr\noalign{\smallskip}
&$\left ]150,\,190\right ]$ &  0.041\,$\pm$\,0.002$^{+0.003}_{-0.004}$ & \cr\noalign{\smallskip}
&$\left ]190,\,230\right ]$ &  0.0291\,$\pm$\,0.0018$^{+0.0033}_{-0.0023}$ & \cr\noalign{\smallskip}
&$\left ]230,\,270\right ]$ &  0.0153\,$\pm$\,0.0015$^{+0.0024}_{-0.0028}$ & \cr\noalign{\smallskip}
\hline\hline\noalign{\smallskip}
\end{tabular*}
}
{\footnotesize
\begin{tabular*}{10.0cm}{@{\extracolsep{0pt}}ll@{\extracolsep{\fill}}r@{\extracolsep{0pt}}l}
\noalign{\smallskip}&\begin{minipage}{2.7cm}$p_\text{T}\,[\text{GeV}]$\hfill\end{minipage} & $\text{d}\sigma_\text{vis}(e^+p\rightarrow e^+D^{\ast\pm} X)/\text{d}p_\text{T}\,[\text{nb}\,\text{GeV}^{-1}]$ & \cr\noalign{\smallskip}\hline\hline\noalign{\smallskip}
&$\left [1.5,\,2\right ]$ &  3.3\,$\pm$\,0.3$^{+0.5}_{-0.3}$ & \cr\noalign{\smallskip}
&$\left ]2,\,2.5\right ]$ &  3.16\,$\pm$\,0.20$^{+0.23}_{-0.31}$ & \cr\noalign{\smallskip}
&$\left ]2.5,\,3.5\right ]$ &  1.92\,$\pm$\,0.08$^{+0.13}_{-0.17}$ & \cr\noalign{\smallskip}
&$\left ]3.5,\,5\right ]$ &  0.79\,$\pm$\,0.04$^{+0.06}_{-0.07}$ & \cr\noalign{\smallskip}
&$\left ]5,\,10\right ]$ &  0.104\,$\pm$\,0.007$^{+0.010}_{-0.009}$ & \cr\noalign{\smallskip}
\hline\hline\noalign{\smallskip}
\end{tabular*}
}
{\footnotesize
\begin{tabular*}{10.0cm}{@{\extracolsep{0pt}}ll@{\extracolsep{\fill}}r@{\extracolsep{0pt}}l}
\noalign{\smallskip}&\begin{minipage}{2.7cm}$\eta$\hfill\end{minipage} & $\text{d}\sigma_\text{vis}(e^+p\rightarrow e^+D^{\ast\pm} X)/\text{d}\eta\,[\text{nb}]$ & \cr\noalign{\smallskip}\hline\hline\noalign{\smallskip}
&$\left [-1.5,\,-1\right ]$ &  2.15\,$\pm$\,0.14$^{+0.20}_{-0.21}$ & \cr\noalign{\smallskip}
&$\left ]-1,\,-0.5\right ]$ &  2.43\,$\pm$\,0.13$^{+0.17}_{-0.20}$ & \cr\noalign{\smallskip}
&$\left ]-0.5,\,0\right ]$ &  2.48\,$\pm$\,0.14$^{+0.17}_{-0.19}$ & \cr\noalign{\smallskip}
&$\left ]0,\,0.5\right ]$ &  2.56\,$\pm$\,0.15$^{+0.21}_{-0.26}$ & \cr\noalign{\smallskip}
&$\left ]0.5,\,1\right ]$ &  2.49\,$\pm$\,0.15$^{+0.28}_{-0.23}$ & \cr\noalign{\smallskip}
&$\left ]1,\,1.5\right ]$ &  1.86\,$\pm$\,0.16$^{+0.18}_{-0.23}$ & \cr\noalign{\smallskip}
\hline\hline\noalign{\smallskip}
\end{tabular*}
}
 \end{center}
\vspace*{-4mm} \caption{\label{txsec1}{
Differential cross sections for inclusive $D^{*\pm}$ meson production in bins of $Q^2$, $x$, $W$, $p_{\rm T}$ and $\eta$. The first error is statistical and the second is systematic.}} 
\end{table}
}
\clearpage

{\large
\begin{table}
 \begin{center}
\vspace*{-6mm}
{\footnotesize
\begin{tabular*}{10.0cm}{@{\extracolsep{0pt}}ll@{\extracolsep{\fill}}r@{\extracolsep{0pt}}l}
\noalign{\smallskip}&\begin{minipage}{2.7cm}$z$\hfill\end{minipage} & $\text{d}\sigma_\text{vis}(e^+p\rightarrow e^+D^{\ast\pm} X)/\text{d}z\,[\text{nb}]$ & \cr\noalign{\smallskip}\hline\hline\noalign{\smallskip}
&$\left [0,\,0.1\right ]$ &  5.9\,$\pm$\,0.8$^{+1.2}_{-0.8}$ & \cr\noalign{\smallskip}
&$\left ]0.1,\,0.2\right ]$ &  12.2\,$\pm$\, 0.8$^{+ 1.4}_{- 1.3}$ & \cr\noalign{\smallskip}
&$\left ]0.2,\,0.3\right ]$ &  11.9\,$\pm$\, 0.7$^{+ 1.4}_{- 1.3}$ & \cr\noalign{\smallskip}
&$\left ]0.3,\,0.4\right ]$ &  9.4\,$\pm$\,0.6$^{+0.9}_{-1.0}$ & \cr\noalign{\smallskip}
&$\left ]0.4,\,0.5\right ]$ &  11.1\,$\pm$\, 0.6$^{+ 0.8}_{- 1.0}$ & \cr\noalign{\smallskip}
&$\left ]0.5,\,0.7\right ]$ &  8.0\,$\pm$\,0.4$^{+0.8}_{-1.5}$ & \cr\noalign{\smallskip}
&$\left ]0.7,\,1\right ]$ &  1.35\,$\pm$\,0.11$^{+0.41}_{-0.52}$ & \cr\noalign{\smallskip}
\hline\hline\noalign{\smallskip}
\end{tabular*}
}
 \end{center}
\vspace*{-3mm} \caption{\label{txsec2}{
Differential cross sections for inclusive $D^{*\pm}$ meson production
in bins of $z$. The first error is statistical and the second is systematic. }}
\end{table}

\vspace*{8mm}
\begin{table}
 \begin{center}
{\footnotesize
\begin{tabular*}{15.0cm}{@{\extracolsep{0pt}}ll@{\extracolsep{\fill}}lr@{\extracolsep{0pt}}l}
\noalign{\smallskip}&\begin{minipage}{2.7cm}$Q^2\,[\text{GeV}^2]$\hfill\end{minipage} & \begin{minipage}{2.7cm}$x$\hfill\end{minipage} & $\text{d}\sigma_\text{vis}(e^+p\rightarrow e^+D^{\ast\pm} X)/\text{d}Q^2\text{d}x\,[\text{nb}\,\text{GeV}^{-2}]$ & \cr\noalign{\smallskip}\hline\hline\noalign{\smallskip}
&$\left [2,\,4.22 \right]$&$\left [2.51 \cdot 10^{\text{-5}},\,5.01 \cdot 10^{\text{-5}}\right ]$ & 2315\,$\pm$\, 525$^{+ 368}_{- 556}$ & \cr\noalign{\smallskip}
&&$\left ]5.01 \cdot 10^{\text{-5}},\,0.0001\right ]$ & 5432\,$\pm$\, 495$^{+ 422}_{- 635}$ & \cr\noalign{\smallskip}
&&$\left ]0.0001,\,0.000158\right ]$ & 4160\,$\pm$\, 359$^{+ 320}_{- 365}$ & \cr\noalign{\smallskip}
&&$\left ]0.000158,\,0.000251\right ]$ & 2154\,$\pm$\, 199$^{+ 204}_{- 355}$ & \cr\noalign{\smallskip}
&&$\left ]0.000251,\,0.000501\right ]$ & 741\,$\pm$\, 85$^{+ 99}_{- 72}$ & \cr\noalign{\smallskip}
\hline\noalign{\smallskip}
&$\left ]4.22,\,10 \right]$&$\left [0.0001,\,0.000158\right ]$ & 899\,$\pm$\,116$^{+141}_{-131}$ & \cr\noalign{\smallskip}
&&$\left ]0.000158,\,0.000251\right ]$ & 757\,$\pm$\, 80$^{+ 81}_{- 76}$ & \cr\noalign{\smallskip}
&&$\left ]0.000251,\,0.000501\right ]$ & 413\,$\pm$\, 33$^{+ 31}_{- 42}$ & \cr\noalign{\smallskip}
&&$\left ]0.000501,\,0.001\right ]$ & 162\,$\pm$\, 15$^{+ 15}_{- 15}$ & \cr\noalign{\smallskip}
\hline\noalign{\smallskip}
&$\left ]10,\,17.8 \right]$&$\left [0.000251,\,0.000501\right ]$ & 206\,$\pm$\, 20$^{+ 19}_{- 18}$ & \cr\noalign{\smallskip}
&&$\left ]0.000501,\,0.001\right ]$ & 77\,$\pm$\, 9$^{+ 7}_{- 9}$ & \cr\noalign{\smallskip}
&&$\left ]0.001,\,0.01\right ]$ & 4.9\,$\pm$\,0.5$^{+0.5}_{-0.7}$ & \cr\noalign{\smallskip}
\hline\noalign{\smallskip}
&$\left ]17.8,\,31.6 \right]$&$\left [0.000251,\,0.000501\right ]$ & 34\,$\pm$\, 7$^{+ 3}_{- 4}$ & \cr\noalign{\smallskip}
&&$\left ]0.000501,\,0.001\right ]$ & 46\,$\pm$\, 4$^{+ 7}_{- 6}$ & \cr\noalign{\smallskip}
&&$\left ]0.001,\,0.01\right ]$ & 3.5\,$\pm$\,0.4$^{+0.3}_{-0.4}$ & \cr\noalign{\smallskip}
\hline\noalign{\smallskip}
&$\left ]31.6,\,100 \right]$&$\left [0.001,\,0.00251\right ]$ & 4.1\,$\pm$\,0.4$^{+0.8}_{-0.4}$ & \cr\noalign{\smallskip}
&&$\left ]0.00251,\,0.01\right ]$ & 0.59\,$\pm$\,0.07$^{+0.05}_{-0.08}$ & \cr\noalign{\smallskip}
\hline\hline\noalign{\smallskip}
\end{tabular*}
}
 \end{center}
\vspace*{-3mm} \caption{\label{txsec22}{
Double differential cross sections for inclusive $D^{*\pm}$ meson production in bins of $Q^2$ and $x$. The first error is statistical and the second is systematic. }}
\end{table}
\vspace*{-3mm}
}
\clearpage

\vspace*{8mm}
\begin{table}
 \begin{center}
\vspace*{-6mm}
{\footnotesize
\begin{tabular*}{10.0cm}{@{\extracolsep{0pt}}ll@{\extracolsep{\fill}}r@{\extracolsep{0pt}}l}
\noalign{\smallskip}&\begin{minipage}{2.7cm}$Q^2\,[\text{GeV}^2]$\hfill\end{minipage} & $\text{d}\sigma_\text{vis}(e^+p\rightarrow e^+D^{\ast\pm} X)/\text{d}Q^2\,[\text{nb}\,\text{GeV}^{-2}]$ & \cr\noalign{\smallskip}\hline\hline\noalign{\smallskip}
&$\left [2,\,4.22\right ]$ &  0.54\,$\pm$\,0.03$^{+0.04}_{-0.05}$ & \cr\noalign{\smallskip}
&$\left ]4.22,\,10\right ]$ &  0.181\,$\pm$\,0.010$^{+0.021}_{-0.019}$ & \cr\noalign{\smallskip}
&$\left ]10,\,17.8\right ]$ &  0.069\,$\pm$\,0.005$^{+0.005}_{-0.008}$ & \cr\noalign{\smallskip}
&$\left ]17.8,\,31.7\right ]$ &  0.031\,$\pm$\,0.003$^{+0.003}_{-0.004}$ & \cr\noalign{\smallskip}
&$\left ]31.7,\,100\right ]$ &  0.0058\,$\pm$\,0.0005$^{+0.0010}_{-0.0009}$ & \cr\noalign{\smallskip}
\hline\hline\noalign{\smallskip}
\end{tabular*}
}
{\footnotesize
\begin{tabular*}{10.0cm}{@{\extracolsep{0pt}}ll@{\extracolsep{\fill}}r@{\extracolsep{0pt}}l}
\noalign{\smallskip}&\begin{minipage}{2.7cm}$x$\hfill\end{minipage} & $\text{d}\sigma_\text{vis}(e^+p\rightarrow e^+D^{\ast\pm} X)/\text{d}x\,[\text{nb}]$ & \cr\noalign{\smallskip}\hline\hline\noalign{\smallskip}
&$\left [2.8 \cdot 10^{\text{-5}},\,0.0002\right ]$ &  7916\,$\pm$\, 407$^{+ 606}_{- 791}$ & \cr\noalign{\smallskip}
&$\left ]0.0002,\,0.0005\right ]$ &  3340\,$\pm$\, 186$^{+ 250}_{- 316}$ & \cr\noalign{\smallskip}
&$\left ]0.0005,\,0.0013\right ]$ &  974\,$\pm$\, 60$^{+107}_{-104}$ & \cr\noalign{\smallskip}
&$\left ]0.0013,\,0.0032\right ]$ &  205\,$\pm$\, 18$^{+ 26}_{- 43}$ & \cr\noalign{\smallskip}
&$\left ]0.0032,\,0.02\right ]$ &  7.1\,$\pm$\,1.1$^{+0.5}_{-1.9}$ & \cr\noalign{\smallskip}
\hline\hline\noalign{\smallskip}
\end{tabular*}
}
{\footnotesize
\begin{tabular*}{10.0cm}{@{\extracolsep{0pt}}ll@{\extracolsep{\fill}}r@{\extracolsep{0pt}}l}
\noalign{\smallskip}&\begin{minipage}{2.7cm}$p_\text{T}\,[\text{GeV}]$\hfill\end{minipage} & $\text{d}\sigma_\text{vis}(e^+p\rightarrow e^+D^{\ast\pm} X)/\text{d}p_\text{T}\,[\text{nb}\,\text{GeV}^{-1}]$ & \cr\noalign{\smallskip}\hline\hline\noalign{\smallskip}
&$\left [1.5,\,2\right ]$ &  0.64\,$\pm$\,0.11$^{+0.21}_{-0.05}$ & \cr\noalign{\smallskip}
&$\left ]2,\,2.5\right ]$ &  1.47\,$\pm$\,0.12$^{+0.11}_{-0.18}$ & \cr\noalign{\smallskip}
&$\left ]2.5,\,3.5\right ]$ &  1.20\,$\pm$\,0.06$^{+0.09}_{-0.14}$ & \cr\noalign{\smallskip}
&$\left ]3.5,\,5\right ]$ &  0.65\,$\pm$\,0.03$^{+0.05}_{-0.07}$ & \cr\noalign{\smallskip}
&$\left ]5,\,10\right ]$ &  0.093\,$\pm$\,0.007$^{+0.009}_{-0.009}$ & \cr\noalign{\smallskip}
\hline\hline\noalign{\smallskip}
\end{tabular*}
}
{\footnotesize
\begin{tabular*}{10.0cm}{@{\extracolsep{0pt}}ll@{\extracolsep{\fill}}r@{\extracolsep{0pt}}l}
\noalign{\smallskip}&\begin{minipage}{2.7cm}$\eta$\hfill\end{minipage} & $\text{d}\sigma_\text{vis}(e^+p\rightarrow e^+D^{\ast\pm} X)/\text{d}\eta\,[\text{nb}]$ & \cr\noalign{\smallskip}\hline\hline\noalign{\smallskip}
&$\left [-1.5,\,-1\right ]$ &  0.92\,$\pm$\,0.08$^{+0.06}_{-0.13}$ & \cr\noalign{\smallskip}
&$\left ]-1,\,-0.5\right ]$ &  1.18\,$\pm$\,0.08$^{+0.08}_{-0.11}$ & \cr\noalign{\smallskip}
&$\left ]-0.5,\,0\right ]$ &  1.27\,$\pm$\,0.09$^{+0.09}_{-0.14}$ & \cr\noalign{\smallskip}
&$\left ]0,\,0.5\right ]$ &  1.40\,$\pm$\,0.09$^{+0.11}_{-0.16}$ & \cr\noalign{\smallskip}
&$\left ]0.5,\,1\right ]$ &  1.42\,$\pm$\,0.09$^{+0.21}_{-0.21}$ & \cr\noalign{\smallskip}
&$\left ]1,\,1.5\right ]$ &  1.03\,$\pm$\,0.10$^{+0.10}_{-0.15}$ & \cr\noalign{\smallskip}
\hline\hline\noalign{\smallskip}
\end{tabular*}
}
 \end{center}
\vspace*{-3mm} \caption{\label{txsec3}{
Differential cross sections for inclusive $D^{*\pm}$ meson production for $p_{\rm T}^* > 2.0$ GeV in bins of $Q^2$, $x$, $p_{\rm T}$ and 
$\eta$. The first error is statistical and the second is systematic. }}
\end{table}
\clearpage

{\large
\vspace*{1mm}
\begin{table}
 \begin{center}
{\footnotesize
\begin{tabular*}{10.0cm}{@{\extracolsep{0pt}}ll@{\extracolsep{\fill}}r@{\extracolsep{0pt}}l}
\noalign{\smallskip}&\begin{minipage}{2.7cm}$Q^2\,[\text{GeV}^2]$\hfill\end{minipage} & $\text{d}\sigma_\text{vis}(e^+p\rightarrow e^+D^{\ast\pm} \text{jj}X)/\text{d}Q^2\,[\text{nb}\,\text{GeV}^{-2}]$ & \cr\noalign{\smallskip}\hline\hline\noalign{\smallskip}
&$\left [2,\,4.22\right ]$ &  0.197\,$\pm$\,0.021$^{+0.027}_{-0.016}$ & \cr\noalign{\smallskip}
&$\left ]4.22,\,10\right ]$ &  0.059\,$\pm$\,0.007$^{+0.012}_{-0.008}$ & \cr\noalign{\smallskip}
&$\left ]10,\,17.8\right ]$ &  0.028\,$\pm$\,0.004$^{+0.004}_{-0.004}$ & \cr\noalign{\smallskip}
&$\left ]17.8,\,31.7\right ]$ &  0.0154\,$\pm$\,0.0022$^{+0.0014}_{-0.0029}$ & \cr\noalign{\smallskip}
&$\left ]31.7,\,100\right ]$ &  0.0033\,$\pm$\,0.0005$^{+0.0008}_{-0.0008}$ & \cr\noalign{\smallskip}
\hline\hline\noalign{\smallskip}
\end{tabular*}
}
{\footnotesize
\begin{tabular*}{10.0cm}{@{\extracolsep{0pt}}ll@{\extracolsep{\fill}}r@{\extracolsep{0pt}}l}
\noalign{\smallskip}&\begin{minipage}{2.7cm}$x$\hfill\end{minipage} & $\text{d}\sigma_\text{vis}(e^+p\rightarrow e^+D^{\ast\pm} \text{jj}X)/\text{d}x\,[\text{nb}]$ & \cr\noalign{\smallskip}\hline\hline\noalign{\smallskip}
&$\left [2.8 \cdot 10^{\text{-5}},\,0.0002\right ]$ &  2688\,$\pm$\, 288$^{+ 345}_{- 253}$ & \cr\noalign{\smallskip}
&$\left ]0.0002,\,0.0005\right ]$ &  1255\,$\pm$\, 140$^{+ 112}_{- 114}$ & \cr\noalign{\smallskip}
&$\left ]0.0005,\,0.0013\right ]$ &  438\,$\pm$\, 49$^{+ 61}_{- 71}$ & \cr\noalign{\smallskip}
&$\left ]0.0013,\,0.02\right ]$ &  13.0\,$\pm$\, 1.9$^{+ 1.5}_{- 2.7}$ & \cr\noalign{\smallskip}
\hline\hline\noalign{\smallskip}
\end{tabular*}
}
{\footnotesize
\begin{tabular*}{10.0cm}{@{\extracolsep{0pt}}ll@{\extracolsep{\fill}}r@{\extracolsep{0pt}}l}
\noalign{\smallskip}&\begin{minipage}{2.7cm}$E^\text{max}_\text{T}\,[\text{GeV}]$\hfill\end{minipage} & $\text{d}\sigma_\text{vis}(e^+p\rightarrow e^+D^{\ast\pm} \text{jj}X)/\text{d}E^\text{max}_\text{T}\,[\text{nb}\,\text{GeV}^{-1}]$ & \cr\noalign{\smallskip}\hline\hline\noalign{\smallskip}
&$\left [4,\,6.5\right ]$ &  0.32\,$\pm$\,0.02$^{+0.02}_{-0.03}$ & \cr\noalign{\smallskip}
&$\left ]6.5,\,10\right ]$ &  0.138\,$\pm$\,0.014$^{+0.035}_{-0.013}$ & \cr\noalign{\smallskip}
&$\left ]10,\,20\right ]$ &  0.018\,$\pm$\,0.004$^{+0.010}_{-0.005}$ & \cr\noalign{\smallskip}
\hline\hline\noalign{\smallskip}
\end{tabular*}
}
{\footnotesize
\begin{tabular*}{10.0cm}{@{\extracolsep{0pt}}ll@{\extracolsep{\fill}}r@{\extracolsep{0pt}}l}
\noalign{\smallskip}&\begin{minipage}{2.7cm}$M_\text{jj}\,[\text{GeV}]$\hfill\end{minipage} & $\text{d}\sigma_\text{vis}(e^+p\rightarrow e^+D^{\ast\pm} \text{jj}X)/\text{d}M_\text{jj}\,[\text{nb}\,\text{GeV}^{-1}]$ & \cr\noalign{\smallskip}\hline\hline\noalign{\smallskip}
&$\left [0,\,11\right ]$ &  0.058\,$\pm$\,0.005$^{+0.005}_{-0.004}$ & \cr\noalign{\smallskip}
&$\left ]11,\,17\right ]$ &  0.089\,$\pm$\,0.008$^{+0.019}_{-0.011}$ & \cr\noalign{\smallskip}
&$\left ]17,\,25\right ]$ &  0.026\,$\pm$\,0.005$^{+0.002}_{-0.004}$ & \cr\noalign{\smallskip}
&$\left ]25,\,50\right ]$ &  0.0023\,$\pm$\,0.0012$^{+0.0018}_{-0.0008}$ & \cr\noalign{\smallskip}
\hline\hline\noalign{\smallskip}
\end{tabular*}
}
{\footnotesize
\begin{tabular*}{15.0cm}{@{\extracolsep{0pt}}ll@{\extracolsep{\fill}}lr@{\extracolsep{0pt}}l}
\noalign{\smallskip}&\begin{minipage}{2.7cm}$Q^2\,[\text{GeV}^2]$\hfill\end{minipage} & \begin{minipage}{2.7cm}$\Delta\phi\,[^\circ]$\hfill\end{minipage} & $\text{d}\sigma_\text{vis}(e^+p\rightarrow e^+D^{\ast\pm} \text{jj}X)/\text{d}\Delta\phi\text{d}Q^2\,[\text{nb}\,\text{GeV}^{-2}\,^{\circ-1}]$ & \cr\noalign{\smallskip}\hline\hline\noalign{\smallskip}
&$\left [2,\,10 \right]$&$\left [57.3,\,152\right ]$ & 0.020\,$\pm$\,0.003$^{+0.003}_{-0.002}$ & \cr\noalign{\smallskip}
&&$\left ]152,\,166\right ]$ & 0.122\,$\pm$\,0.017$^{+0.011}_{-0.027}$ & \cr\noalign{\smallskip}
&&$\left ]166,\,180\right ]$ & 0.142\,$\pm$\,0.019$^{+0.037}_{-0.012}$ & \cr\noalign{\smallskip}
\hline\noalign{\smallskip}
&$\left ]10,\,100 \right]$&$\left [57.3,\,132\right ]$ & 0.00079\,$\pm$\,0.00016$^{+0.00008}_{-0.00021}$ & \cr\noalign{\smallskip}
&&$\left ]132,\,166\right ]$ & 0.0054\,$\pm$\,0.0007$^{+0.0025}_{-0.0007}$ & \cr\noalign{\smallskip}
&&$\left ]166,\,180\right ]$ & 0.0124\,$\pm$\,0.0018$^{+0.0010}_{-0.0015}$ & \cr\noalign{\smallskip}
\hline\hline\noalign{\smallskip}
\end{tabular*}
}
 \end{center}
\vspace*{-3mm} \caption{\label{txsec4}{
Differential cross sections for the production of $D^{*\pm}$ mesons with dijets in bins of $Q^2$, $x$, $E_{\rm T}^{\rm max}$, $M_{\rm jj}$ and double differentially in bins of $\Delta\phi$ and $Q^2$. The first error is statistical and the second is systematic. }}
\end{table}
}
\clearpage

{\large
\begin{table}
 \begin{center}
\vspace*{-6mm}
{\footnotesize
\begin{tabular*}{10.0cm}{@{\extracolsep{0pt}}ll@{\extracolsep{\fill}}r@{\extracolsep{0pt}}l}
\noalign{\smallskip}&\begin{minipage}{2.7cm}$\eta_\text{DJ}$\hfill\end{minipage} & $\text{d}\sigma_\text{vis}(e^+p\rightarrow e^+D^{\ast\pm} \text{jj}X)/\text{d}\eta_\text{DJ}\,[\text{nb}]$ & \cr\noalign{\smallskip}\hline\hline\noalign{\smallskip}
&$\left [-2,\,0.9\right ]$ &  0.066\,$\pm$\,0.010$^{+0.007}_{-0.007}$ & \cr\noalign{\smallskip}
&$\left ]0.9,\,1.5\right ]$ &  0.66\,$\pm$\,0.06$^{+0.11}_{-0.07}$ & \cr\noalign{\smallskip}
&$\left ]1.5,\,2\right ]$ &  0.62\,$\pm$\,0.07$^{+0.07}_{-0.10}$ & \cr\noalign{\smallskip}
&$\left ]2,\,3\right ]$ &  0.45\,$\pm$\,0.05$^{+0.11}_{-0.07}$ & \cr\noalign{\smallskip}
&$\left ]3,\,5\right ]$ &  0.058\,$\pm$\,0.014$^{+0.018}_{-0.015}$ & \cr\noalign{\smallskip}
\hline\hline\noalign{\smallskip}
\end{tabular*}
}
{\footnotesize
\begin{tabular*}{10.0cm}{@{\extracolsep{0pt}}ll@{\extracolsep{\fill}}r@{\extracolsep{0pt}}l}
\noalign{\smallskip}&\begin{minipage}{2.7cm}$\eta_\text{OJ}$\hfill\end{minipage} & $\text{d}\sigma_\text{vis}(e^+p\rightarrow e^+D^{\ast\pm} \text{jj}X)/\text{d}\eta_\text{OJ}\,[\text{nb}]$ & \cr\noalign{\smallskip}\hline\hline\noalign{\smallskip}
&$\left [-2,\,0.9\right ]$ &  0.048\,$\pm$\,0.010$^{+0.009}_{-0.005}$ & \cr\noalign{\smallskip}
&$\left ]0.9,\,1.5\right ]$ &  0.43\,$\pm$\,0.06$^{+0.04}_{-0.06}$ & \cr\noalign{\smallskip}
&$\left ]1.5,\,2\right ]$ &  0.65\,$\pm$\,0.08$^{+0.09}_{-0.07}$ & \cr\noalign{\smallskip}
&$\left ]2,\,3\right ]$ &  0.48\,$\pm$\,0.05$^{+0.13}_{-0.05}$ & \cr\noalign{\smallskip}
&$\left ]3,\,5\right ]$ &  0.117\,$\pm$\,0.019$^{+0.010}_{-0.021}$ & \cr\noalign{\smallskip}
\hline\hline\noalign{\smallskip}
\end{tabular*}
}
{\footnotesize
\begin{tabular*}{10.0cm}{@{\extracolsep{0pt}}ll@{\extracolsep{\fill}}r@{\extracolsep{0pt}}l}
\noalign{\smallskip}&\begin{minipage}{2.7cm}$|\Delta \eta|$\hfill\end{minipage} & $\text{d}\sigma_\text{vis}(e^+p\rightarrow e^+D^{\ast\pm} \text{jj}X)/\text{d}|\Delta \eta|\,[\text{nb}]$ & \cr\noalign{\smallskip}\hline\hline\noalign{\smallskip}
&$\left [0,\,0.3\right ]$ &  0.90\,$\pm$\,0.12$^{+0.32}_{-0.10}$ & \cr\noalign{\smallskip}
&$\left ]0.3,\,0.8\right ]$ &  0.82\,$\pm$\,0.09$^{+0.06}_{-0.07}$ & \cr\noalign{\smallskip}
&$\left ]0.8,\,1.4\right ]$ &  0.64\,$\pm$\,0.07$^{+0.14}_{-0.09}$ & \cr\noalign{\smallskip}
&$\left ]1.4,\,2\right ]$ &  0.36\,$\pm$\,0.05$^{+0.03}_{-0.04}$ & \cr\noalign{\smallskip}
&$\left ]2,\,4\right ]$ &  0.070\,$\pm$\,0.014$^{+0.024}_{-0.006}$ & \cr\noalign{\smallskip}
\hline\hline\noalign{\smallskip}
\end{tabular*}
}
 \end{center}
\caption{\label{txsec5}{
Differential cross sections for the production of $D^{*}$-jet and other jet in bins of $\eta_{\rm DJ}$, $\eta_{\rm OJ}$ and 
$\Delta\eta$. The first error is statistical and the second is systematic. }}
\end{table}

\vspace*{-6mm}
\begin{table}
 \begin{center}
{\footnotesize
\begin{tabular*}{10.0cm}{@{\extracolsep{0pt}}ll@{\extracolsep{\fill}}r@{\extracolsep{0pt}}l}
\noalign{\smallskip}&\begin{minipage}{2.7cm}$x^\text{obs}_\gamma$\hfill\end{minipage} & $\text{d}\sigma_\text{vis}(e^+p\rightarrow e^+D^{\ast\pm} \text{jj}X)/\text{d}x^\text{obs}_\gamma\,[\text{nb}]$ & \cr\noalign{\smallskip}\hline\hline\noalign{\smallskip}
&$\left [0.2,\,0.45\right ]$ &  0.59\,$\pm$\,0.14$^{+0.39}_{-0.15}$ & \cr\noalign{\smallskip}
&$\left ]0.45,\,0.7\right ]$ &  0.94\,$\pm$\,0.16$^{+0.13}_{-0.09}$ & \cr\noalign{\smallskip}
&$\left ]0.7,\,1\right ]$ &  3.5\,$\pm$\,0.2$^{+0.4}_{-0.3}$ & \cr\noalign{\smallskip}
\hline\hline\noalign{\smallskip}
\end{tabular*}
}
{\footnotesize
\begin{tabular*}{15.0cm}{@{\extracolsep{0pt}}ll@{\extracolsep{\fill}}lr@{\extracolsep{0pt}}l}
\noalign{\smallskip}&\begin{minipage}{2.7cm}$Q^2\,[\text{GeV}^2]$\hfill\end{minipage} & \begin{minipage}{2.7cm}$x^\text{obs}_\gamma$\hfill\end{minipage} & $\text{d}\sigma_\text{vis}(e^+p\rightarrow e^+D^{\ast\pm} \text{jj}X)/\text{d}x^\text{obs}_\gamma\text{d}Q^2\,[\text{nb}\,\text{GeV}^{-2}]$ & \cr\noalign{\smallskip}\hline\hline\noalign{\smallskip}
&$\left [2,\,5 \right]$&$\left [0.2,\,0.7\right ]$ & 0.112\,$\pm$\,0.022$^{+0.036}_{-0.013}$ & \cr\noalign{\smallskip}
&&$\left ]0.7,\,1\right ]$ & 0.35\,$\pm$\,0.04$^{+0.05}_{-0.04}$ & \cr\noalign{\smallskip}
\hline\noalign{\smallskip}
&$\left ]5,\,10 \right]$&$\left [0.2,\,0.7\right ]$ & 0.015\,$\pm$\,0.009$^{+0.014}_{-0.006}$ & \cr\noalign{\smallskip}
&&$\left ]0.7,\,1\right ]$ & 0.146\,$\pm$\,0.019$^{+0.039}_{-0.015}$ & \cr\noalign{\smallskip}
\hline\noalign{\smallskip}
&$\left ]10,\,100 \right]$&$\left [0.2,\,0.7\right ]$ & 0.0034\,$\pm$\,0.0007$^{+0.0017}_{-0.0005}$ & \cr\noalign{\smallskip}
&&$\left ]0.7,\,1\right ]$ & 0.0184\,$\pm$\,0.0017$^{+0.0014}_{-0.0020}$ & \cr\noalign{\smallskip}
\hline\hline\noalign{\smallskip}
\end{tabular*}
}
 \end{center}
\vspace*{-3mm}
\caption{\label{txsec6}{
Differential cross sections for the production of $D^{*}$-jet and other jet in bins of $x_{\gamma}^{\rm obs}$ and double differentially in three ranges of $Q^2$. The first error is statistical and the second is systematic. }}
\end{table}
\vspace*{-5mm}
\begin{table}
 \begin{center}
{\footnotesize
\begin{tabular*}{10.0cm}{@{\extracolsep{0pt}}ll@{\extracolsep{\fill}}r@{\extracolsep{0pt}}l}
\noalign{\smallskip}&\begin{minipage}{2.7cm}$\log_{10} x^\text{obs}_\text{g}$\hfill\end{minipage} & $\text{d}\sigma_\text{vis}(e^+p\rightarrow e^+D^{\ast\pm} \text{jj}X)/\text{d}x^\text{obs}_\text{g}\,[\text{nb}]$ & \cr\noalign{\smallskip}\hline\hline\noalign{\smallskip}
&$\left [-3.3,\,-2.4\right ]$ &  38\,$\pm$\, 7$^{+ 4}_{- 4}$ & \cr\noalign{\smallskip}
&$\left ]-2.4,\,-2.1\right ]$ &  108\,$\pm$\, 10$^{+  9}_{- 10}$ & \cr\noalign{\smallskip}
&$\left ]-2.1,\,-1.8\right ]$ &  61\,$\pm$\, 5$^{+ 7}_{- 6}$ & \cr\noalign{\smallskip}
&$\left ]-1.8,\,-0.9\right ]$ &  3.6\,$\pm$\,0.5$^{+1.8}_{-0.7}$ & \cr\noalign{\smallskip}
\hline\hline\noalign{\smallskip}
\end{tabular*}
}
{\footnotesize
\begin{tabular*}{15.0cm}{@{\extracolsep{0pt}}ll@{\extracolsep{\fill}}lr@{\extracolsep{0pt}}l}
\noalign{\smallskip}&\begin{minipage}{2.7cm}$Q^2\,[\text{GeV}^2]$\hfill\end{minipage} & \begin{minipage}{2.7cm}$\log_{10} x^\text{obs}_\text{g}$\hfill\end{minipage} & $\text{d}\sigma_\text{vis}(e^+p\rightarrow e^+D^{\ast\pm} \text{jj}X)/\text{d}x^\text{obs}_\text{g}\text{d}Q^2\,[\text{nb}\,\text{GeV}^{-2}]$ & \cr\noalign{\smallskip}\hline\hline\noalign{\smallskip}
&$\left [2,\,5 \right]$&$\left [-3.3,\,-2.4\right ]$ & 5.2\,$\pm$\,1.4$^{+1.9}_{-0.4}$ & \cr\noalign{\smallskip}
&&$\left ]-2.4,\,-2.1\right ]$ & 12.0\,$\pm$\, 2.0$^{+ 1.2}_{- 1.3}$ & \cr\noalign{\smallskip}
&&$\left ]-2.1,\,-1.8\right ]$ & 5.8\,$\pm$\,1.1$^{+0.8}_{-0.8}$ & \cr\noalign{\smallskip}
&&$\left ]-1.8,\,-0.9\right ]$ & 0.36\,$\pm$\,0.10$^{+0.27}_{-0.05}$ & \cr\noalign{\smallskip}
\hline\noalign{\smallskip}
&$\left ]5,\,10 \right]$&$\left [-3.3,\,-2.4\right ]$ & 1.14\,$\pm$\,0.55$^{+0.28}_{-0.15}$ & \cr\noalign{\smallskip}
&&$\left ]-2.4,\,-2.1\right ]$ & 5.0\,$\pm$\,1.0$^{+1.2}_{-1.0}$ & \cr\noalign{\smallskip}
&&$\left ]-2.1,\,-1.8\right ]$ & 2.4\,$\pm$\,0.5$^{+0.5}_{-0.3}$ & \cr\noalign{\smallskip}
&&$\left ]-1.8,\,-0.9\right ]$ & 0.178\,$\pm$\,0.050$^{+0.016}_{-0.064}$ & \cr\noalign{\smallskip}
\hline\noalign{\smallskip}
&$\left ]10,\,100 \right]$&$\left [-3.3,\,-2.4\right ]$ & 0.192\,$\pm$\,0.055$^{+0.018}_{-0.023}$ & \cr\noalign{\smallskip}
&&$\left ]-2.4,\,-2.1\right ]$ & 0.49\,$\pm$\,0.07$^{+0.04}_{-0.04}$ & \cr\noalign{\smallskip}
&&$\left ]-2.1,\,-1.8\right ]$ & 0.34\,$\pm$\,0.04$^{+0.06}_{-0.04}$ & \cr\noalign{\smallskip}
&&$\left ]-1.8,\,-0.9\right ]$ & 0.018\,$\pm$\,0.004$^{+0.005}_{-0.004}$ & \cr\noalign{\smallskip}
\hline\hline\noalign{\smallskip}
\end{tabular*}
}
 \end{center}
\vspace*{-3mm}
\caption{\label{txsec7}{
Differential cross sections for the production of $D^{*}$-jet and other jet in bins of $x_{\rm g}^{\rm obs}$ and double differentially in three ranges of $Q^2$. The first error is statistical and the second is systematic. }}
\end{table}
\vspace*{-5mm}
}

\clearpage

\clearpage

\clearpage 

\clearpage

\clearpage

\clearpage

\clearpage

\clearpage

\end{document}